\documentclass{article}
\usepackage{amsmath}
\usepackage{graphicx}

\begin{document}
\title{Bound and resonant impurity states in a narrow gaped armchair
graphene nanoribbon.}
\author{B.~S.~Monozon\\
Physics Department, Marine Technical University, 3 Lotsmanskaya Str.,\\
190008 St.Petersburg, Russia,\\
\\
P.~Schmelcher\\
Zentrum f\"ur Optische Quantentechnologien, Universit\"{a}t Hamburg, \\ Luruper Chaussee 149, 22761 Hamburg, Germany}
\date{\today}
\maketitle

\begin{abstract}
An analytical study of discrete and resonant impurity quasi-Coulomb states in a narrow gaped
armchair graphene nanoribbon (GNR) is performed. We employ the adiabatic approximation
assuming that the motions parallel ("slow") and perpendicular ("fast") to the boundaries
of the ribbon are separated adiabatically. The energy spectrum comprises a sequence
of series of quasi-Rydberg levels relevant to the "slow" motion adjacent from the low
energies to the size-quantized levels associated with the "fast" motion. Only the series
attributed to the ground size-quantized sub-band is really discrete, while others
corresponding to the excited sub-bands consist of quasi-discrete (Fano resonant) levels
of non-zero energetic widths, caused by the coupling with the states of the continuous
spectrum branching from the low lying sub-bands. In the two- and three-subband approximation
the spectrum of the complex energies of the impurity electron is derived in an explicit
form. Narrowing the GNR leads to an increase of the binding energy and the resonant width
both induced by the finite width of the ribbon. Displacing the impurity centre from the
mid-point of the GNR causes the binding energy to decrease while the resonant width of the
first excited Rydberg series increases. As for the second excited series their widths
become narrower
with the shift of the impurity. A successful
comparison of our analytical results with those obtained by other theoretical
and experimental methods is presented.
Estimates of the binding energies and the resonant widths taken for the parameters
of typical GNRs show that not only the strictly discrete but also the some
resonant states are quite
stable and could be studied experimentally in doped GNRs.

\end{abstract}

\section{Introduction}\label{S:intro}

The electron properties of two-dimensional (2D) graphene,
a single-layer carbon sheet, has attracted much attention by both theoreticians and experimentalists
(see \cite{Castro} and references therein). Along with this related
structures, namely graphene nanoribbons, are also under intensive investigation \cite{Rosl}.
One of the reason for this is that the long electron mean free path in graphene up to
1\,$\mu$m opens a field of carbon-based nanoelectronics, where GNRs are used as
interconnects in nanodevices. The unique electron mobility in graphene structures
is caused by the strong bonding between the carbon atoms in the honeycomb lattice of graphene.
This in turn prevents the replacing of the carbon atoms by alien ones. Nevertheless,
graphene is not immune to extrinsic disorder and its transport properties \cite{Namura}
are very sensitive to impurities and defects \cite{Peres}.

The theoretical problem
of an impurity in 2D graphene was considered originally in \cite{Bis,Nov,Per,Shyt,Shyt1}.
In the vicinity of the Dirac points in $\vec{k}$ space, which are peaks of the double cones
of the Fermi surface, the low-energy electronic excitations
in gapless graphene are described by the equation
of the effective mass approximation, which is formally identical to the
2D Dirac equation for a massless neutrino, having the Fermi speed $v_F=10^6\,\mbox{m/c}$.
In the presence of an attractive impurity centre of charge $Z$ screened by a medium of
the effective dielectric constant $\epsilon_{\mbox{eff}}$ the electron states are drastically different
for the subcritical $J_c < J$ and supercritical $J_c > J$ regions of the
strength of the Coulomb interaction, where $J = |j|\hbar,\,|j|=1,2,\ldots $
and $J_c = Ze^2/4\pi\epsilon_0\epsilon_{\mbox{eff}} v_F$ are the 2D
momentum of the impurity electron and that,
having the speed $v_F$, respectively. Clearly, the super- and subcritical
regimes can be reached
for the dimensionless Coulomb potential strength
$q=Ze^2/4\pi\varepsilon_0\epsilon_{\mbox{eff}} \hbar v_F$
for $q>|j|~\mbox{and}~q<|j|$, respectively.
The difference of the subcritical and supercritical electron states is caused by
their different behaviour in the vicinity of the impurity centre $\vec{r}\longrightarrow 0$.
The subcritical regime admits regular solutions to the Dirac equations, while the
wave functions corresponding to the supercritical case oscillate and do not have
any definite limit. The physical reason for this is that at the subcritical strengths
$g<|j|$ the centrifugal potential barrier prevents the electron "fall to the centre"
\cite{landau}, while the supercritical strengths $g>|j|$ provide the collapse.
Clearly, the continuum approach based on the Dirac formalism becomes inapplicable.
The lattice-scale physics dominates that in turn requires a regularization procedure,
namely the cutoff of the Coulomb potential at short distances $r_0 \simeq a$, where
$a\simeq 1.42~ {\rm {\AA}}$ is the C-C distance in graphene. The physics of the supercritical
impurity electron in graphene \cite{Nov} closely resembles that
of the relativistic electron in an atom having the nuclear charge
$Z>137$ \cite{perpop,pop70,pop71,zeld}. Since, as it follows from below,
only the supercritical regime
is relevant to the impurity state in GNR, we focus on this case.

Numerical and analytical approaches developed on the tight-binding model of
the graphene lattice and of the Dirac equation, subject to the regularization procedure,
respectively, undertaken originally by Pereira
\cite{Per} et.al. have revealed the infinite number of the quasi-bound states,
having the finite width, arising, as it was shown quasi-classically \cite{Shyt1},
from the collapsed states.
If the requirement of the regularity of the wave
functions in the vicinity of the source of the electron attraction is to be replaced by the less
rigorous condition of its square integrability an infinite number
of the strictly discrete energy levels were found to occur.
The Coulomb potential cutoff in the gapped \cite{gupta}
and no cutoff in the gapless graphene \cite{gupta1} induce
the energy series bounded and unbounded from below, respectively.

In the GNR, which in principle can be treated as a quasi-1D structure, we can expect
completely different results. The strictly discrete bound states
regular at the impurity centre $(\vec{r}=0)$ are realized without
the regularization procedure,
in particular without the cutoff of the Coulomb potential, preventing the collapse.
Consequential concerns lie in the well known fact that
the reduction
of the dimension of the structure increases the stability of the impurity electron. In units of the
impurity Rydberg constant $Ry$ the binding energy $E_b$ of the impurity electron in 3D bulk material
is $E_b =1$, in the narrow 2D quantum well $E_b = 4$ \cite{harr} , and in the thin quantum wire of radius
$R$ much less than the impurity Bohr radius $a_0$, $E_b \sim \ln^2 (R/a_0)$ \cite{monschm09}.
Besides, an extremely weak 3D atomic potential not providing bound electron states,
transforms in the presence of a magnetic field into a quasi-1D system binding
the electron \cite{demdruk}. It is relevant to note that these atomic states arise under
as weak as one likes magnetic fields i.e. the as large as one likes magnetic lengths
playing the same role as the width of the GNR. Note that the confinement
attributed to the semiconductor thin films \cite{keld} replaces the
3D Coulomb potential $(\sim r^{-1})$
by the effective 2D potential of the weaker singularity of the logarithmic character.
It seems that in the quasi-1D GNR
the effect of the attenuation of the potential singularity preventing the
fall to the centre exceeds that of the vanishing of the 2D centrifugal potential
barrier promoting the collapse.

Clearly a study of the impurity electron state in graphene structures is important on account
of two reasons. First, these structures
provide a realization in solid state physics of remarkable effects of quantum electrodynamics
caused by a large "fine structure constant" $e^2/\hbar v_F \simeq 2.5$ \cite{Shyt,Shyt1,katsn,katsnnov}.
Second, we expect a strong impact of impurities
on the electronic systems not only for 2D graphene layers
possessing an outstanding high electron mobility \cite{Nov} but in particular
for impurity GNRs whose properties are not
widely addressed in the literature yet.

Brey and Fertig \cite{Brey} have shown that the energy spectrum of the electron
in an armchair GNR bounded in $x$-direction is the sequence of the subbands
formed by the branches of
 the continuous energies of the longitudinal
unbounded $y$-motion emanating from the size-quantized energy levels
$\varepsilon_N $, ($N$ is the discrete lable),
reflecting the ribbons $x$-confinement. The equation for the components $u_{A,B} (y)$ of
the Dirac spinor relevant to the A and B sublattices of the graphene for the
electron positioned far away from the impurity centre has the form

$$
u_{A,B}^{\prime \prime}(y)+\frac{E^2 - \varepsilon_N ^2}{\hbar^2 v_F^2}\,u_{A,B}(y)=0,
$$
showing that the armchair GNR manifests itself as gapped structure entailing the bound and unbound
impurity states for the energies $E^2 <\varepsilon_N ^2~\mbox{and}~E^2 \geq \varepsilon_N ^2$, respectively.

Of special interest is the narrow GNR of width $d$ for which $d << r_0$,
where $r_0^2(d)\simeq \hbar^2 v_F^2(E^2 - \varepsilon_N ^2)^{-1}$
is the radius of the bound electron state,
being induced by the ribbon confinement $d$.
Such a GNR provides the expected electron binding energy $E_b \sim \varepsilon_N f(q)$,
where $f(q)$ is a some function vanishing at $q=0$,
which is of interest and attractive because of two aspects.
On the one hand the ribbon provides
a considerable impurity binding energy which could be measured experimentally
and on the other hand the impurity potential can be treated perturbatively
and an analytical approach
to the problem becomes feasible.

\begin{figure}[t!]
\begin{center}
\includegraphics*[width=.7\columnwidth]{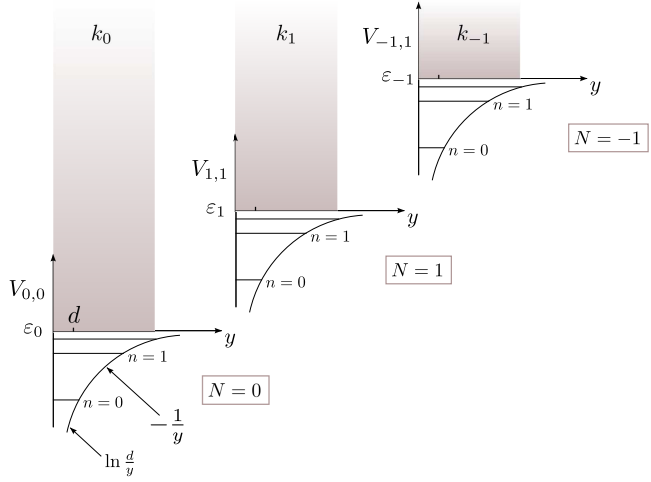}
\end{center}
\caption{\label{fig1} A schematic form of the potentials $V_{NN}(y)$
provided in eqs. (\ref{E:coulomb}),
(\ref{E:pot}), (\ref{E:limit}) at $x_0=0$ and quasi-discrete $n$ (\ref{E:rydberg})
and continuous $k_N$ (\ref{E:outercont}) spectra adjacent to the ground $N=0$ (discrete states)
and first $N=1$ and second $N=-1$ size-quantized levels $\varepsilon_N$ (\ref{E:phi})
in the GNR of width $d$.}
\end{figure}

A comment concerning the from of the energy spectrum is in order. In the zeroth approximation
of isolated size-quantized $N$-subbands i.e. in the single-subband approximation the slow
longitudinal motion parallel to the boundaries is governed by the 2D Coulomb potential
averaged with respect to the "fast" transverse $N$-states.
The energy spectrum consists of sequence of series of quasi-Coulomb
discrete $Nn$-levels and continuous sub-bands positioned below and above, respectively
relatively to the
size-quantized energy levels $\varepsilon_N$ (see Fig. 1).
Only the series of the impurity energy levels $E_{0n}<\varepsilon_0$
adjacent to the ground size-quantized energy level $\varepsilon_0 $ is strictly discrete.
 The $Nn$-series adjacent to the excited
levels $N>0$ come into resonance with the states of the continuous spectra of lower subbands
and in fact in the next multi-subband approximation turn into quasi-discrete resonant states
(Fano resonances)\cite{fano}. The corresponding resonant widths $\Gamma_{Nn}$
determine the auto-ionization rate and
life-time $\tau_{Nn}=\hbar/\Gamma_{Nn}$ of the resonant impurity states
 being of relevance to an
experimental study. Also to our knowledge an analytical approach to the problem of impurities
in GNR providing the explicit dependencies of both discrete and especially quasi-discrete
electron states on the width of the GNR $(d)$ and the position of the impurity centre
within the GNR are not comprehensively available in the literature.

In order to fill the above mentioned gap we perform an analytical study of the strictly discrete
and resonant impurity states in a narrow armchair GNR. The impurity
centre is positioned anywhere within the ribbon bound by the impenetrable boundaries.
The width of the GNR is assumed to be much less than the radius of the impurity state.
The complete 2D envelope wave function satisfying
the massless Dirac equation is expanded with respect to the basis formed by the 1D
size-quantized subband wave functions describing the fast transverse motion bound by the
boundaries of the GNR. The generated set of equations for the 1D quasi-Coulomb wave functions
relevant to the longitudinal slow motion is solved in the single-, two- and three-subband
approximations,
in which the ground, first and second excited subbands are involved. The mathematical method is based
on the matching of the Coulomb wave functions with those obtained by an iteration procedure at
any point within the intermediate region bound by the ribbon width and the radius of the
created Coulomb state. Both the real and imaginary parts of the complex energy levels are
calculated in a single procedure. The dependencies of the binding energy
and resonant energy shift and width on the width of the GNR and the position of the impurity centre
are obtained in an explicit form. Numerical estimates show that for a narrow GNR the binding energy
and the resonant width are quite reasonable, to render the impurity
electron states in GNR experimentally observable. Our analytical results are in
line with those calculated numerically and revealed in an experiment.
We remark that our aim is to elucidate the physics
of the impurity states in GNR by deriving closed form analytical expressions for their properties.
We do not intend to compete with the results of computational studies.

The paper is organized as follows. In Section 2 the analytical approach based on the
multi-subband approximation is described. The real quasi-Coulomb functions of the discrete
and continuous spectrum and the real energy levels determining the binding energies
are calculated in the single-subband approximation in Section 3. The complex energies including
the resonant shift and width associated to the first and second excited subbands are found
in the double and three-subband approximation, respectively, in Section 4. In Section 5 we discuss
the obtained results. Section 6 contains the conclusions.

According to the above an analytical description
of the stable and metastable impurity electron states in the narrow armchair GNR is of significant
interest. It elucidates the underlying basic physics of the carbon-based nanodevices,
in which the highly mobile electrons remain unbound in the 2D graphene monolayers  while in their interconnects,
namely in quasi-1D armchair GNRs these electrons are trapped by impurity centres. The latter
could modify the overall transport properties.

\section{General approach}\label{S:gen}

We consider a ribbon of width $d$ located in the $x-y$ plane and bounded by the lines $x=\pm d/2.$
The impurity centre of charge $Z$ is displaced from the mid-point of the ribbon
$x=0$ by the distance $-d/2 \leq x_0 \leq d/2.$
The equation describing the impurity electron at a position $\vec{\rho}=(x,y)$
possesses the form of a Dirac equation

\begin{equation}\label{E:basic}
\hat{{\rm H}}(\hat{\vec{k}},\vec{\rho })\vec{\Psi}(\vec{\rho })=E\vec{\Psi}(\vec{\rho });\qquad \hat{\vec{k}}=-i\vec{\nabla};
\end{equation}
where the Hamiltonian $\hat{{\rm H}}$ is given by

$$
\hat{{\rm H}}= p\left[\hat{{\rm H}}_0(\hat{k}_x)+ \hat{{\rm H}}_1(\hat{k}_y) \right]
 +V(\vec{\rho})\hat{{\rm I}};\,p=\hbar v;\,v=10^6~\mbox{m/c}
$$
with

$$
\hat{{\rm H}}_0(\hat{k}_x)= \left(\begin{array}{cc} -\sigma_x \hat{k}_x
& 0 \\0 & \sigma_x \hat{k}_x \\
\end{array} \right) ~;
\qquad \hat{{\rm H}}_1(\hat{k}_y)= \left (\begin{array}{cc} -\sigma_y
\hat{k}_y & 0 \\0 & - \sigma_y \hat{k}_y \end{array} \right)~;
$$
composed by the Hamiltonians relevant to the inequivalent Dirac points
 $\vec{K}^{(+)}~(-\vec{\sigma}\vec{k})$
and $\vec{K}^{(-)}~(-\vec{\sigma}^{*}\vec{k})$ ($\vec{\sigma}=(\sigma_x , \sigma_y)$
 are the Pauli matrixes)
presented originally in Ref. \cite{Brey}. The matrix $\hat{{\rm I}}$ in (\ref{E:basic})
 is the unit matrix and

\begin{equation}\label{E:coulomb}
V(\vec{\rho})=-\frac{\beta}{\sqrt{(x-x_0)^2 + y^2}}~;\qquad \beta = \frac{Ze^2}{4\pi\epsilon_0 \epsilon_{\mbox{eff}}}~;
\end{equation}
is the 2D Coulomb impurity potential, $\epsilon_{\mbox{eff}}$ is the effective dielectric constant related
to the static dielectric constant $\epsilon$ of the substrate by \cite{Nov,hwang}

$$
\epsilon_{\mbox{eff}} =\frac{1}{2}(1+\epsilon +\pi q_0);~q_0 = \frac{e^2}{4\pi\epsilon_0\hbar v_F}\simeq2.2.
$$

The envelope wave four-vector $\vec{\Psi}$ consists of two vectors $\vec{\Psi}_{A,B}$
 describing the motion of the electron in sublattices $A$ and $B$ of graphene

\begin{equation}\label{E:wavevect}
\vec{\Psi}(\vec{\rho})= \vec{\Psi}_A(\vec{\rho})+ \vec{\Psi}_B(\vec{\rho})~;
\end{equation}
each determined by the wave functions $\psi_{A,B}^{(+,-)}$

$$
\vec{\Psi}_A =
\begin{array}{c}
\begin{Bmatrix}
\psi_{A}^{(+)} \\
0 \\
\psi_{A}^{(-)}\\
0
\end{Bmatrix}
\end{array}
~;\qquad
\vec{\Psi}_B =
\begin{array}{c}
\begin{Bmatrix}
 0\\
\psi_{B}^{(+)} \\
0 \\
\psi_{B}^{(-)} \\
\end{Bmatrix}
\end{array}
$$

The total $A,B$ state implies the multiplication of the $\psi_{A,B}^{(+,-)}$
functions with the factors $\exp \{{\rm i} \vec{K}^{(+,-)}\vec{\rho} \}$, respectively. The boundary conditions
for the armchair ribbon require the total wave function to vanish at both edges $x=\pm d/2$ for both
 $A,B$ superlattices \cite{Castro}

\begin{equation}\label{E:bound}
 e^{{\rm i}Kx}\psi_{A,B}^{(+)}(\vec{\rho})+ e^{-{\rm i}Kx}\psi_{A,B}^{(-)}(\vec{\rho})=0 ~\mbox{at}\, x=\pm\frac{d}{2},
\end{equation}
where $\vec{K}^{(+,-)}=(\pm K, 0)~;\, K=4\pi/3a_0,~a_0 = 2.46 ~{\rm {\AA}}$ is the graphene superlattice
constant.

The basis wave vectors $\vec{\Phi}_N (x)$ and the energies $\varepsilon_N$ describing the transverse size-quantized
$x$-states are derived from equation

\begin{equation}\label{E:trans}
\hat{{\rm H}}_0(\hat{k}_x)\vec{\Phi}_N (x)=\varepsilon_N \vec{\Phi}(x)
\end{equation}
to obtain

$$
\vec{\Phi}_N (x)=\frac{1}{\sqrt{2}}\left [ \vec{\Phi}_{NA} (x) + \vec{\Phi}_{NB} (x)  \right ];~
\vec{\Phi}_{NA} (x)=
\begin{array}{c}
\begin{Bmatrix}
\varphi_{NA}^{(+)} \\
0 \\
\varphi_{NA}^{(-)}\\
0
\end{Bmatrix}
\end{array}
;~ \vec{\Phi}_{NB} =
\begin{array}{c}
\begin{Bmatrix}
 0\\
\varphi_{NB}^{(+)} \\
0 \\
\varphi_{NB}^{(-)}
\end{Bmatrix}
\end{array}
$$,
where

\begin{equation}\label{E:sizef}
-\varphi_{NA}^{(+)}=\varphi_{NA}^{(-)*}=\varphi_{NB}^{(+)}=-\varphi_{NB}^{(-)*}=\varphi_{N0}
\end{equation}
with

\begin{eqnarray}\label{E:phi}
\varphi_{N0}(x)=\frac{1}{\sqrt{2d}}\exp \left \{ {\rm i}\left [x\frac{\pi}{d}(N-\tilde{\sigma})-\frac{\pi}{2}
\left(N+ \left[ \frac{Kd}{\pi} \right]    \right)  \right ]\right \};
\nonumber\\
\varepsilon_N=|N-\tilde{\sigma}|\frac{\pi p}{d};~N=0,\pm1,\pm2,\ldots~;
\quad\tilde{\sigma}=\frac{Kd}{\pi}-\left[ \frac{Kd}{\pi}\right]
\end{eqnarray}

We consider transverse states with positive energies $\varepsilon_N > 0$ in the armchair ribbon
of width $d$ providing the gaped (insulator) structure $\tilde{\sigma} \neq 0$ \cite{Castro}.
It follows from eq.(\ref{E:phi}) that the energy levels $\varepsilon_N$ as a function
of width $d$ are the oscillations describing by parameter $\tilde{\sigma}(d)$ imposed
on the decreasing curve $\sim d^{-1}$. Below we ignore these oscillations keeping
$\tilde{\sigma}=\mbox{const.}$

 The boundary conditions (\ref{E:bound})
after substitution of $\psi_{A,B}^{(+,-)}$
by $\varphi_{NA,B}^{(+,-)}$, respectively, are satisfied. The wave vectors $\vec{\Phi}_{NA,B}$
 form orthonormal subsets, for which

\begin{eqnarray}\label{E:ortho}
\hat{{\rm H}}_0(\hat{k}_x) \vec{\Phi}_{NA(B)}&=&\varepsilon_N \vec{\Phi}_{NB(A)}~;\,
\langle\vec{\Phi}_{N'A(B)}|\vec{\Phi}_{NB(A)}\rangle =0~;
\nonumber\\
\langle\vec{\Phi}_{N'A(B)}|\vec{\Phi}_{NA(B)}\rangle &=& \delta_{N'N}~;
\qquad \langle\vec{\Phi}_{N'}|\vec{\Phi}_{N}\rangle =\delta_{N'N}.
 \end{eqnarray}

The boundary conditions (\ref{E:bound}) imposed on the wave vector $\vec{\Psi}(\vec{\rho})$ (\ref{E:wavevect})
force us to expand the wave vectors $\vec{\Psi}_{A,B}$ in series

\begin{equation}\label{E:expan}
\vec{\Psi}_{A,B}(\vec{\rho})=\Sigma_N u_{NA,B}(y)\vec{\Phi}_{NA,B}(x)~; \,\mbox{i.e.}\,
\psi_{A,B}^{(+,-)}(\vec{\rho})=\Sigma_N u_{NA,B}(y)\varphi_{NA,B}^{(+,-)}(x)~
\end{equation}
with respect to the basis functions $\varphi_{NA,B}^{(+,-)}(x)$ taking for the coefficients
$u_{NA,B}^{(+)}(y)=u_{NA,B}^{(-)}(y)\equiv u_{NA,B}(y).$
Substituting the wave vector $\vec{\Psi}$ (\ref{E:wavevect}) with the wave vectors $\vec{\Psi}_{A,B}$ and the
wave functions $\psi_{A,B}^{(+,-)}(\vec{\rho})$ into eq. (\ref{E:basic})
and subsequently using the properties (\ref{E:ortho}) we obtain by the standard method the set of equations
for the wave functions

$$
v_N^{(1)}=\frac{1}{\sqrt{2}}(u_{NB}+u_{NA});\quad v_N^{(2)}=\frac{1}{\sqrt{2}}(u_{NB}-u_{NA})~;
$$

\begin{eqnarray}\label{E:set}
\left.
\begin{array}{c}
\frac{dv_N^{(1)}(y)}{dy} - \frac{1}{p}\left(E + \varepsilon_N -V_{NN}(y)\right)v_N^{(2)}(y)
+\frac{1}{p}\sum_{N'\neq N} V_{N'N}(y)v_{N'}^{(2)}(y)=0~;\\
\frac{dv_N^{(2)}(y)}{dy} + \frac{1}{p}\left(E - \varepsilon_N -V_{NN}(y)\right)v_N^{(1)}(y)
-\frac{1}{p}\sum_{N'\neq N} V_{N'N}(y)v_{N'}^{(1)}(y)=0~;
\end{array}
\right \}
\end{eqnarray}

\begin{equation}\label{E:pot}
V_{N'N}(y)=\frac{1}{d}\int_{-\frac{d}{2}}^{+\frac{d}{2}} V(\vec{\rho})
\cos \left[(N-N')\pi \left(\frac{x}{d}- \frac{1}{2}\right)dx  \right],
\end{equation}
where the potential $V(\vec{\rho})$ is given by eq. (\ref{E:coulomb}).
At $|y|\gg d$

\begin{equation}\label{E:onoff}
V_{N'N}(y)=-\frac{\beta}{|y|}\left[ \delta_{N'N}+O \left(\frac{d^2}{y^2}  \right)
\delta_{|N'-N|(2s+1)} \right]~;\,s=0,1,2,\ldots ;
\end{equation}

As expected in the limiting case $d\rightarrow 0$ eqs. (\ref{E:set}) decompose into the sets describing
the 1D Coulomb states, while in the absence of the impurity centre $(V_{N'N}=0)$ we arrive at the wave
functions $u_{NA,B}\sim \exp (\pm {\rm i} k_y y)$ and the energies $E_N^2 (k_y)=\varepsilon_N^2 +p^2 k_y^2$
of free electrons in the armchair nanoribbon \cite{Brey, Castro}.

Below we solve the set (\ref{E:set}) in the adiabatic approximation implying the longitudinal $y$-motion
governed by the quasi-Coulomb potentials $V_{N'N}(y)$ to be much slower than the transverse $x$-motion
affected by the boundaries of the narrow ribbon. The Coulomb potential (\ref{E:coulomb})
 is assumed to be small compared to the ribbon confinement. In the case of $\tilde{\sigma}<0.5$
 the lowest three sub-bands are
specified by indices $N=0,1,-1$. The set (\ref{E:set}) corresponding to these subbands becomes

\begin{eqnarray}\label{E:set1}
\left.
\begin{array}{c}
v_0^{(1)'} - \frac{1}{p}\left(E + \varepsilon_0 -V_{00}\right)v_0^{(2)}
+\frac{1}{p}\left[ V_{10}v_{1}^{(2)}+ V_{-10}v_{-1}^{(2)}\right]=0~;\\
v_0^{(2)'} + \frac{1}{p}\left(E - \varepsilon_0 -V_{00}\right)v_0^{(1)}
-\frac{1}{p} \left[ V_{10}v_{1}^{(1)}+ V_{-10}v_{-1}^{(1)}\right]=0~;\\
v_1^{(1)'} - \frac{1}{p}\left(E + \varepsilon_1 -V_{11}\right)v_1^{(2)}
+\frac{1}{p}\left[ V_{01}v_{0}^{(2)}+ V_{-11}v_{-1}^{(2)}\right]=0~;\\
v_1^{(2)'} + \frac{1}{p}\left(E - \varepsilon_1 -V_{11}\right)v_1^{(1)}
-\frac{1}{p} \left[ V_{01}v_{0}^{(1)}+ V_{-11}v_{-1}^{(1)}\right]=0~;\\
v_{-1}^{(1)'} - \frac{1}{p}\left(E + \varepsilon_{-1} -V_{-1-1}\right)v_{-1}^{(2)}
+\frac{1}{p}\left[ V_{0-1}v_{0}^{(2)}+ V_{1-1}v_{1}^{(2)}\right]=0~;\\
v_{-1}^{(2)'} + \frac{1}{p}\left(E - \varepsilon_{-1} -V_{-1-1}\right)v_{-1}^{(1)}
-\frac{1}{p} \left[ V_{0-1}v_{0}^{(1)}+ V_{1-1}v_{1}^{(1)}\right]=0~;
\end{array}
\right \}
\end{eqnarray}

\section{Single-subband approximation}\label{S:single}

At the first stage we neglect the coupling between the states corresponding to the
subbands with different $N.$ The reason for this is that in the narrow ribbon
of small width $d$ the diagonal potentials $V_{NN}$ dominate the off-diagonal terms
$V_{N'N}~(N'\neq N)$ almost everywhere but for a small region $|y|\leq d$ (see eq. (\ref{E:onoff})).
In this case $V_{N'N}=V_{NN}\delta_{N'N}$ and the set (\ref{E:set}) decomposes into
independent subsets each specified by an index $N$. The 1D impurity states
are then governed by the potential

\begin{eqnarray}\label{E:limit}
V_{NN}(y)=\frac{\beta}{d}\ln\frac{\frac{4y^2}{d_1 d_2}}{\left(1 + \sqrt{1+\frac{4y^2}{d_1 ^2}} \right)
\left(1 + \sqrt{1+\frac{4y^2}{d_2
^2}} \right)}=
\left\{
\begin{array}{cl}
\frac{\beta}{d}\ln\frac{y^2}{d_1 d_2}~;\, &\frac{|y|}{d_{1,2}}\ll 1\\
-\frac{\beta}{|y|}~;\, &\frac{|y|}{d_{1,2}}\gg 1
\end{array}
\right.
\end{eqnarray}

$$
d_{1,2}=d\pm 2x_0~;\qquad -\frac{d}{2} <x_0 <+\frac{d}{2}~;
$$

The set (\ref{E:set}) for $y>0$ with $V_{N'N}=0$
is solved by matching in the intermediate region the
corresponding solutions $\{v_N^{(1)}, v_N^{(2)}\}$ one of which is valid in the inner region close to the
impurity centre and the other represents a solution of the outer region distant from the centre.
This method was
originally developed by Hasegava and Howard \cite{hashow} in studies of excitons subject to
strong magnetic fields and then successfully employed for the investigation of the impurity and exciton
states in quantum wells \cite{monschm05}, super-lattices \cite{monschm07} and
quantum wires \cite{monschm09}.

\subsection{Inner region}

In the inner region \\
 $0\leq y \ll r_0~(r_0 = p|\varepsilon_N ^2 - E^2|^{-1/2}\,
\mbox{is the effective size
of the Coulomb state})$ an iteration procedure is performed. The first integration
of the set (\ref{E:set}), in which we neglect the terms consisting of the energies
$\varepsilon_N~\mbox{and}~E$, with the trial functions
$v_{N0}^{(1)}=a_N^{(1)};v_{N0}^{(2)}=a_N^{(2)}$  gives

\begin{equation}\label{E:step1}
v_{N1}^{(1)}(y)=a_N^{(1)} - a_N^{(2)}\frac{q}{2d}\left[d_1 F(f_1)+ d_2 F(f_2) \right]~;\quad f_{1,2}=\frac{2y}{d_{1,2}}
\end{equation}
where

$$
F(f)=f\ln\frac{\sqrt{1+f^2}-1}{f}- \mbox{arsh} f=
\left\{
\begin{array}{cl}
f\left(\ln\frac{|f|}{2}-1 \right)~,\, &|f|\ll 1\\
-\frac{f}{|f|}(\ln 2|f| +1)~,\,\,& |f|\gg 1
\end{array}
\right.
$$
and where

$$
q=\frac{\beta}{p}=\frac{Ze^2}{4\pi\epsilon_0\epsilon_{\mbox{eff}}\hbar v_F};~~(q\ll 1)
$$
is the dimensionless strength of the Coulomb potential.

The function $v_{N1}^{(2)}(y)$ can be obtained from eq. (\ref{E:step1}) by replacing $a_N^{(1)}$ by $a_N^{(2)}$
and $a_N^{(2)}$ by $-a_N^{(1)}$. Subsequent integration leads to the two independent particular solutions
 $\{v_{N+}^{(1)},v_{N+}^{(2)}\}$ and $\{v_{N-}^{(1)}, v_{N-}^{(2)}\}$ corresponding to the
relationships $a_N^{(2)}=\pm {\rm i}a_N^{(1)}.$ The linear combination of these solutions
provides the general iteration functions, which read in the region $y \gg d_{1,2}$

\begin{equation}\label{E:iter}
v_{N \mbox{in} }^{(1)}(y)=R_N\sin(Q+\zeta_N)~; v_{N \mbox{in}}^{(2)}(y)=R_N\cos(Q+\zeta_N)~;
\,Q(y)=q\frac{y}{|y|}\left( \ln\frac{4|y|}{D}+1 \right),
\end{equation}
where

\begin{equation}\label{E:widthef}
D=\sqrt{d_1 d_2}\exp \left\{ \frac{1}{4d}(d_1 - d_2)\ln\frac{d_1}{d_2}\right\}
\end{equation}
and where $R_N$ and $\zeta_N$ are the arbitrary magnitude and phase, respectively.
Since the potentials (\ref{E:pot}) satisfy $V_{NN'}(y)$ = $V_{NN'}(-y)$
the wave two-vectors\\
$\vec{v}_{N}\{v_N^{(1)}, v_N^{(2)} \}$ are
classified with respect to parity. Further we focus on the even wave vectors
for which $\hat{\Pi}\vec{v}_{N}=\vec{v}_{N}$ where $\hat{\Pi}=\hat{\pi}\sigma_z$
with $\hat{\pi}v(y)=v(-y)$ ($\sigma_z$ is the Pauli matrix). The condition of even
parity imposed on the wave vector $\vec{v}_{N \mbox{in} }$ formed by the components
(\ref{E:iter}), implies that the phases $\zeta_N$ are equal to the half integer
of $\pi$.

Obviously, an alternative way to derive eq.
(\ref{E:iter}) is to solve eqs. (\ref{E:set}) for $V_{NN}(y) =-\beta |y|^{-1},~\varepsilon_N =E=0$ and then
to compare the resulting solutions $v_{N+,-}^{(1,2)}(y)\sim A y^{\pm {\rm i}q}$, expanded in series up to the terms
of the first order of $q$, with those given by eqs. (\ref{E:step1}). The calculated constant $A$
leads to the functions (\ref{E:iter}).

\subsection{Outer region}

a) \emph{Discrete states}

The exact solutions to eqs. (\ref{E:set}) at $V_{N'N}=0$ for $N\neq N'$
in the region $y\gg d_{1,2}$ with
$V_{NN}(y)=-\beta y^{-1}$ are calculated by the same method employed in studies of a
relativistic electron in hydrogen \cite{berlif} and super-heavy atoms with the
nuclear charge number
number $Z>137$ \cite{perpop,pop70,pop71,zeld}

\begin{eqnarray}\label{E:outer}
\begin{Bmatrix}
v^{(1)}_{N}(\tau) \\
v^{(2)}_{N}(\tau) \\
\end{Bmatrix}
=A_N
\left\{
\begin{array}{ll}
\cosh\frac{\psi_N}{2}\tau^{-\frac{1}{2}}\left[ W_{\kappa,\mu}(\tau)
+\frac{\tanh \psi}{q}W_{\kappa +1,\mu}(\tau) \right]\\
\sinh\frac{\psi_N}{2}\tau^{-\frac{1}{2}}\left[ W_{\kappa,\mu}(\tau)
-\frac{\tanh \psi}{q}W_{\kappa +1,\mu}(\tau) \right]
\end{array}
\right.
\end{eqnarray}
where

\begin{eqnarray}
\tau&=&\nu_N y,~\nu_N = \frac{2}{p}\sqrt{\varepsilon_N^2-E^2};\,\tanh \psi =\frac{p\nu_N}{2\varepsilon_N};~
\kappa = \eta-\frac{1}{2};~\mu ={\rm i q};
\nonumber\\
\eta_N&=&\frac{2qE}{p\nu_N};~
A_N^2= \frac{\nu_N}{2\Gamma(\eta_N)^2\cosh\psi_N\left(1+\frac{\eta_N^2}{q^2}\tanh\psi_N^2 \right)}
(1+\delta_{\eta\eta_0});
\nonumber\\
\eta_0 &\ll& 1~\mbox{is the quantum number labeling the ground state}
\nonumber
\end{eqnarray}
and where $W_{\kappa,\mu}(\tau)$ is the Whittaker function associated with the Kummer
function $U$ \cite{abram}

\begin{equation}\label{E:whitt}
W_{\kappa,\mu}(\tau) = e^{-\frac{\tau}{2}}\tau^{\frac{1}{2}+\mu}U(a,c,\tau);\,a=\frac{1}{2}+\mu - \kappa,\,
c=1+2\mu,
\end{equation}
with \cite{baterd}

\begin{eqnarray}\label{E:kum}
U(a,c,\tau)=
\left\{
\begin{array}{ll}
\tau^{-a}; & \tau \gg 1 \\
\frac{\Gamma (1-c)}{\Gamma (a-c+1)}+\frac{\Gamma (c-1)}{\Gamma (a)}\tau^{1-c}; \mbox{Re}~ c =1;\, c\neq 1;\, & \tau\ll 1.
\end{array}
\right.
\end{eqnarray}
The functions (\ref{E:outer}) are normalized to $\int_{-\infty}^{+\infty}(v_N^{(1)2} + v_N^{(2)2})dy=1$.

The asymptotic behavior of the outer functions (\ref{E:outer}) at large distances $y\gg r_0$
follows from eqs. (\ref{E:outer}),(\ref{E:whitt}) and (\ref{E:kum})

\begin{equation}\label{E:asimpt}
v_{N}^{(1,2)}(y) \sim \exp\left( -\frac{y}{r_0}+\eta_N\ln\frac{2y}{r_0} \right);
\,r_0=\frac{2}{\nu_N}.
\end{equation}

In the region $\tau \ll 1$ eqs. (\ref{E:outer}),(\ref{E:whitt}) and (\ref{E:kum}) lead
to the expressions

\begin{equation}\label{E:outer1}
v_{N \mbox{out}}^{(1)}=P_N\sin\omega_N~; v_{N \mbox{out}}^{(2)}=P_N\cos\omega_N~;
\,\omega_N (\tau)=q\ln\tau +\Theta_N,
\end{equation}
where

\begin{equation}\label{E:theta}
\Theta_N = \arg \Gamma(-2{\rm i}q) + \arg \Gamma(-\eta_N + {\rm i}q) -
\arctan \frac{\eta_N}{q}\left(\sqrt{1+\frac{q^2}{\eta_N^2}}-1  \right),
\end{equation}
and $P_N$ are arbitrary constants.

Matching the eqs. (\ref{E:outer}) and (\ref{E:iter}) in the overlapping intermediate
region $d\ll y \ll r_0$ we impose the condition

$$
\frac{v_{N \mbox{in}}^{(1)}(y)}{v_{N \mbox{in}}^{(2)}(y)} =\frac{v_{N \mbox{out}}^{(1)}(y)}{v_{N \mbox{out}}^{(2)}(y)}
$$
which yields

\begin{equation}\label{E:eqn}
\omega_N -Q -\zeta_N = s\pi;\, s=0,\pm1,\pm2,\ldots .
\end{equation}

Using the properties of the arguments of the $\Gamma$-functions in eq. (\ref{E:theta})
 for a small parameter $q \ll 1$ and for the quantum numbers
$\eta_N = n + \delta_{Nn} , n =0,1,2,\ldots, \delta_{Nn} < 1$ \cite{abram}
 (see also Ref. \cite{perpop,pop71} for details)
and choosing $\zeta_N = \pi/2$ we arrive at the equation for the corrections
$\delta_{Nn}$

\begin{eqnarray}\label{E:discr}
\ln q +  \frac{1}{q} \left[ \arctan\frac{q}{\delta_{Nn}} - \arctan \frac{q}{2(n+\delta_{Nn})} \right]
-\ln (n+\delta_{Nn}) \qquad \qquad
\nonumber \\
+~~ \psi(1+n)+\ln \frac{|N-\tilde{\sigma}|\pi D}{2d}+2C-1=0,
\end{eqnarray}
where $n=0,1,2,\ldots , \psi(1+n)$ is the psi-function
(logarithmic derivative of the $\Gamma$-function), $C=0.577$ is the Euler constant. The
corrections $\delta_{Nn}$ calculated from eq. (\ref{E:discr}) determine the Rydberg series
of the discrete energy levels $E_{Nn}$ adjacent to the size-quantized energy level $\varepsilon_N$

\begin{eqnarray}\label{E:rydberg}
E_{Nn}=
\left\{
\begin{array}{ll}
\varepsilon_N \left[1- \varepsilon_N\frac{q^2}{2(n+\delta_{Nn})^2}\right];\, &n=1,2,\ldots \\
\frac{\varepsilon_N}{\sqrt{1+\frac{q^2}{\delta_{N0}^2}}};~&n=0
\end{array}
\right.
\end{eqnarray}
which allow to estimate the size of the Coulomb state in eq. (\ref{E:asimpt})
$r_0 \simeq d/|N-\tilde{\sigma}|\pi q$
for $n=1,2,\ldots$ and

$$
r_0 =\frac{d}{|N-\tilde{\sigma}|\pi\sqrt{1-\frac{1}{1+\frac{q^2}{\delta_{N0}^2}}} }~\mbox{for}~n=0.
$$
Clearly from (\ref{E:rydberg}), the existence of the intermediate
matching region $d_{1,2} \ll y \ll r_0$ is provided for excited state
$n=1,2,\ldots$ by the employed above small parameter $q\ll 1$
and for the ground state $n=0$ by the condition

$$
(1\pm \frac{2x_0}{d})|N-\tilde{\sigma}|\pi z_0
\ll 1,~\mbox{i.e.}~z_0=\frac{q}{\delta_{N0}}\ll 1.
$$
The correction $\delta_{N0}=q/z_0$ satisfies the transcendental
equation

\begin{equation}\label{E:ground}
\ln z_0 +\frac{1}{q}\left[\arctan z_0-\arctan\frac{z_0}{2}\right]
+\ln \frac{|N-\tilde{\sigma}|\pi D}{2d}+C-1=0,
\end{equation}
while the corrections $\delta_{Nn}$ for the excited states $n=1,2,\ldots$ can be calculated in an
explicit form
\begin{equation}\label{E:exc}
\delta_{Nn}=q\cot\left\{ q \left[  -\ln q +
\ln n + \frac{1}{2n} -\psi(1+n)-\ln \frac{|N-\tilde{\sigma}|\pi D}{2d}-2C+1\right]\right\},
\end{equation}

b) \emph{Continuous states}

Since our approach to determine the wave function of the
continuous states closely resembles that applied above for the wave functions
of the discrete states only the basic points will be given below. Setting in eqs.
(\ref{E:outer}) $\nu_N =-2{\rm i}k$ we obtain

\begin{eqnarray}\label{E:outercont}
\begin{Bmatrix}
v^{(1)}_{N+}(t) \\
v^{(2)}_{N+}(t) \\
\end{Bmatrix}
=B_N
\left\{
\begin{array}{ll}
&\cos\frac{\varphi_N}{2}t^{-\frac{1}{2}}\left[ W_{\tilde{\kappa},\mu}(t)
-{\rm i}\frac{\tan \varphi}{q}W_{\tilde{\kappa}+1,\mu}(t) \right]\\
-{\rm i}&\sin\frac{\varphi_N}{2}t^{-\frac{1}{2}}\left[ W_{\tilde{\kappa},\mu}(t)
+{\rm i}\frac{\tan \varphi}{q}W_{\tilde{\kappa}+1,\mu}(t) \right].
\end{array}
\right.
\end{eqnarray}
where

\begin{eqnarray}
t&=&-2{\rm i}k_N y;~k_N = \frac{1}{p}\sqrt{E^2 - \varepsilon_N^2};\,\tan\varphi =\frac{pk}{\varepsilon_N};\,
\tilde{\kappa} = {\rm i}\frac{q}{\sin\varphi}-\frac{1}{2};
\nonumber\\
\mu &=& {\rm i }q;\, B_N^2=\frac{q^2}{2\pi\tan^2\varphi}\exp\left(-\frac{\pi q}{\sin\varphi}  \right).
\end{eqnarray}
The wave vectors (\ref{E:outercont}) are normalized to $\delta (k-k')$.

At large distances $ky \gg 1$ the wave functions (\ref{E:outercont}), (\ref{E:whitt})
, (\ref{E:kum}) have the asymptotic form of the outgoing waves

$$
v_{N+}^{(1,2)}(y)\sim\exp\left\{{\rm i }k_N y + {\rm i }\frac{q}{\sin\varphi}\ln 2k_N y \right\}.
$$

Further we introduce the real wave functions associated with the standing waves

$$
v_{N \mbox{out}}^{(1,2)}(t)=D_N\left[e^{{\rm i}\Omega_N}v_{N +}^{(1,2)}(t) + e^{-{\rm i}\Omega_N}v_{N -}^{(1,2)}(t) \right],
$$
where the functions $v_{N -}^{(1,2)}(t)= v_{N +}^{(1,2)*}(t)$ have the asymptotic form
of the ingoing waves $\sim\exp\left\{-{\rm i }ky - {\rm i }\frac{q}{\sin\varphi}\ln 2ky \right\}$
and $D_N$ and $\Omega_N$ are the arbitrary magnitude and phase, respectively.

In the region $|t|\ll 1$ eqs. (\ref{E:outercont}), (\ref{E:whitt}) and
(\ref{E:kum}) lead to

\begin{equation}\label{E:outercont1}
v_{N \mbox{out}}^{(1)}(t)=D_N\left[ e^{q\frac{\pi}{2}}\frac{M(\varphi)}{|\Gamma_{+}|}
\cos\left( \Omega_N +q\ln |t|+ \xi_{+} \right) -
e^{-q\frac{\pi}{2}}\frac{M(-\varphi)}{|\Gamma_{-}|}
\cos\left( -\Omega_N +q\ln |t|+ \xi_{-} \right)\right]
\end{equation}
where

$$
\Gamma_{+,-}=\Gamma\left[ - {\rm i }q \left(\frac{1}{\sin\varphi}\pm 1  \right)\right];
\xi_{+,-}=\arg \Gamma (-2{\rm i }q)\mp \arg \Gamma_{+,-};
M(\varphi)=\frac{\sin\frac{\varphi}{2}+\cos\frac{\varphi}{2}}{1+\sin\varphi}.
$$
The wave functions $v_{N \mbox{out}}^{(2)}(t)$ can be obtained from eq. (\ref{E:outercont1})
by replacing $D_N$ by $-D_N$ and $\xi_{+,-}$ by $\xi_{+,-}-\pi/2.$

For $q\ll 1,~\varphi\ll 1$ the wave functions (\ref{E:outercont1}) read

\begin{equation}\label{E:outercont2}
v_{N \mbox{out}}^{(1)}(t)=D_N\left[ \sin\chi_N -c_N\cot\Lambda_N\cos\chi_N \right];
v_{N \mbox{out}}^{(2)}(t)=D_N\left[ \cos\chi_N +c_N\cot\Lambda_N\sin\chi_N \right]
\end{equation}
with

\begin{equation}\label{E:qfirst}
\chi_N=q\ln |t|+\frac{1}{2}(\xi_{-}+\xi_{+});~\Lambda_N=\Omega_N-\frac{1}{2}(\xi_{-}-\xi_{+});~
c_{N} = q\frac{\pi}{2}\left( 1+\coth\frac{q\pi}{\varphi} \right).
\end{equation}

Similar to the case of the discrete states we obtain the equation
for the phase $\Lambda_N$ on equating the ratios $v_N^{(1)}(y)/v_N^{(2)}(y)$
taken for the iteration (\ref{E:iter}) and outer functions (\ref{E:outercont2})

\begin{equation}\label{E:lambda}
\cot\Lambda_N = \frac{1}{c_N}\tan(\chi - Q - \zeta_N).
\end{equation}

Since \cite{abram}

$$
\frac{1}{2}(\xi_{-}-\xi_{+})=\frac{\pi}{2}-\Omega^{(0)}                          ;\quad
\Omega^{(0)}=\sum_{j=1}^{\infty}\left( \frac{q}{j\varphi}-\arctan\frac{q}{j\varphi} \right)
$$
eq. (\ref{E:lambda}) acquires for $\zeta_N=\pi/2$ an explicit form

\begin{equation}\label{E:omega}
\cot\Omega_N =\frac{\frac{\pi}{2}\left( 1+\coth\frac{q\pi}{\varphi}\right)}
{\ln\frac{2}{kD}-\frac{1}{2}\left[ \psi\left(1+{\rm i}\frac{q}{\sin \varphi}+
\right) + \psi\left(1-{\rm i}\frac{q}{\sin \varphi}+
\right)\right]-2C+1}.
\end{equation}
Since at $q\ll 1$ the phase $\Omega^{(0)}=\frac{1}{3}\zeta(3)\frac{q^3}{\varphi^3}$~($\zeta(s)$~
is the Riemann zeta function with $\zeta(3)=1.20$) is the value of the higher
order of smallness $\sim q^3 \ll 1$ we set in eq. (\ref{E:omega}) $\Omega^{(0)} = 0$.

As expected, setting in the functions $v_{N \mbox{out}}^{(1,2)}$ (\ref{E:outercont1})
$k_N=\frac{{\rm i}\nu_N}{2},~(\Gamma_{+}=\Gamma_{-}^{*})$ and then matching these functions with
the iteration functions $v_{N\mbox{it} }^{(1,2)}$ (\ref{E:iter}) we obtain the
equation for the phase $\Omega_N$. Substituting this result into equation
$\cot\Omega_N = {\rm i}$ determining the poles of the
$S$ matrix with $S=\exp(2{\rm i}\Omega_N)$ \cite{landau, berlif} we arrive at eqs. (\ref{E:eqn}), (\ref{E:discr}) for
the discrete energy levels.

\section{Double-subband approximation}\label{S:double}

Below we consider the coupling between the ground $N=0$ and first excited $N=1$
states described by the system of the four upper eqs. (\ref{E:set1}) at
$V_{-1 0}=V_{-1 1}=0.$ Applying the iteration method with the trial functions
$v_{0}^{(1,2)}=a_0^{(1,2)}~\mbox{and}~v_{1}^{(1,2)}=a_1^{(1,2)}$ we arrive at two particular
linear independent four-vectors, having the components $v_{0,1}^{(1,2)}$ calculated
for $a_0^{(2)}={\rm i}a_0^{(1)}, a_1^{(2)}={\rm i}a_1^{(1)}~ \mbox{and}~
a_0^{(2)}=-{\rm i}a_0^{(1)}, a_1^{(2)}=-{\rm i}a_1^{(1)}$.  The linear combination of these vectors
taken for $a_{0,1}^{(2)}=R_{0,1}\exp \left [ {\rm i}(\zeta_{0,1} - \frac{\pi}{2})\right ] $
provides the general expression for the iteration four-vector with the components

\begin{eqnarray}\label{E:iter2}
\left.
\begin{array}{c}
v_{0 it}^{(1)}(y)=R_0\sin (Q + \zeta_0) + R_1 q \gamma_{01}\cos\zeta_1~;\\
v_{0 it}^{(2)}(y)=R_0\cos (Q + \zeta_0) - R_1 q \gamma_{01}\sin\zeta_1~;
\end{array}
\right \}
\end{eqnarray}
where $R_{0,1}~\mbox{and }~ \zeta_{0,1}$ are an arbitrary constant and phase, respectively.
The parameter

\begin{eqnarray}\label{E:couple}
\pi\gamma_{0,1}&=&\cos\alpha_0 \left[  {\rm {Ci}}\left( \frac{\pi}{2} +\alpha_0\right)-
{\rm {Ci}}\left( \frac{\pi}{2} -\alpha_0\right) \right]
\nonumber\\
&+&\sin\alpha_0 \left[  {\rm {Si}}\left( \frac{\pi}{2} +\alpha_0\right)+
{\rm {Si}}\left( \frac{\pi}{2} -\alpha_0\right) \right],~\alpha_0=\frac{\pi x_0}{d}.
\end{eqnarray}
consisting of the integral sine ${\rm {Si}}~ \mbox{and cosine}~ {\rm {Ci}}$ \cite{abram},
describes the coupling induced by the potentials $V_{01}=V_{10}$ (\ref{E:pot}). The functions
$v_{1 \mbox {in}}^{(1)}(y)~ \mbox{and}~v_{1 \mbox {in}}^{(2)}(y)$ can be obtained from the functions
$v_{0 \mbox {in}}^{(1)}(y)~ \mbox{and}~v_{0 \mbox {in}}^{(2)}(y)$ (\ref{E:iter2}) by mutual replacing
$R_0 \leftrightarrow R_1.$

Equating the ratios of the functions of the continuous spectrum\\ (\ref{E:outercont2})
$v_{0 \mbox{out}}^{(1)}(t)/v_{0 \mbox{out}}^{(2)}(t)$ and the iteration functions (\ref{E:iter2})
$v_{0 \mbox{in}}^{(1)}(y)/v_{0 \mbox{in}}^{(2)}(y)$ and then matching the ratios of the functions
of the discrete states (\ref{E:outer1}) $v_{1 \mbox{out}}^{(1)}(\tau)/v_{1 \mbox{out}}^{(2)}(\tau)$
and the iteration functions $v_{1 \mbox{in}}^{(1)}(y)/v_{1\mbox{in} }^{(2)}(y)$ we obtain the set of
equations

\begin{eqnarray}\label{E:setcoupl}
\left.
\begin{array}{l}
R_0\left[\sin (\chi_0 -Q - \zeta_0)- c_0\cot\Lambda_0\cos (\chi_0 -Q -
\zeta_0) \right]+ \qquad \\
\qquad R_1 q \gamma_{01} \left[\cos(\chi_0 - \zeta_1)- c_0\cot\Lambda_0
\sin (\chi_0 - \zeta_1) \right] =0~; \\
\qquad ~~~ R_0 q \gamma_{01}\cos (\omega_1 - \zeta_0) - R_1
\sin(\omega_1 - Q -\zeta_1) =0 ~,
\end{array}
\right \}
\end{eqnarray}
where the functions $\chi_N, Q,\omega_N$ are defined by eqs. (\ref{E:qfirst}), (\ref{E:iter}),
(\ref{E:outer1}), respectively. In the limiting case of negligible coupling the set (\ref{E:setcoupl})
decomposes into two independent equations relevant to the discrete (\ref{E:eqn}), (\ref{E:discr})
and continuous (\ref{E:lambda}) states.

Solving the set (\ref{E:setcoupl}) by the determinantal method we obtain the equation for $\cot\Lambda_0$,
which is then expanded in series with respect to the parameter $q\ll 1.$ Keeping at
$\zeta_N = \pi/2$ the terms of the first order $\sim q$ we arrive at the equation for the phase
$\Omega_0$ in an explicit form

\begin{equation}\label{E:cot}
\cot\Omega_0 =
\frac{c_0 (\omega_1 - Q - \frac{\pi}{2})}{(\omega_1 - Q - \frac{\pi}{2})(Q-\chi_0 + \frac{\pi}{2})
-q^2\gamma_{0,1}^2}.
\end{equation}

Substituting eq. (\ref{E:cot}) into equation \cite{landau,berlif,newton}

\begin{equation}\label{E:image}
\cot \Omega_0 ={\rm i}
\end{equation}
the complex quantum numbers $\eta_1$ introduced in eq. (\ref{E:outer}) can be calculated,
which in turn determine the complex energy levels $E_{1n}$ adjacent to the size-quantized
first excited energy level $\varepsilon_1 = (1-\tilde{\sigma})\pi p/d$

\begin{equation}\label{E:compl}
E_{1n}=\varepsilon_1 - \varepsilon_1\frac{q^2}{2(n+\delta_{1n}^2)}
+W_{1n} - {\rm i }\frac{\Gamma_{1n}}{2};~ n= 0,1,2\ldots ,
\end{equation}
where the second term in the right-hand part is the Rydberg series of the energy levels
associated with the quasi-Coulomb diagonal potential $V_{11}(y)$ (\ref{E:limit}) (no coupling).
The following notation in eq. (\ref{E:compl}) for the resonant shift $W_{1n}$
and resonant width $\Gamma_{1n}$ both induced by the inter-subband $N=0,1$ interaction is used

\begin{equation}\label{E:width}
\Gamma_{1n} = 2\varepsilon_{1}\frac{q^2}{(n+\delta_{1n})^3}G_n (\delta_{1n})
q\gamma_{0,1}^2 B_{0,1}
\end{equation}
and

\begin{equation}\label{E:shift}
W_{1n} =- \varepsilon_{1}\frac{q^2}{(n+\delta_{1n})^3}G_n (\delta_{1n})
q^2 \gamma_{0,1}^2 A_{0,1}.
\end{equation}
In eqs. (\ref{E:width}) and (\ref{E:shift})

$$
A_{0,1} = B_{0,1}^2\left (\ln\frac{k_0 d}{2} + \ln\frac{D}{d} +C -1 \right);
~~B_{0,1}=\frac{2\varepsilon_0}{pk_0}.
$$

$$
G_0^{-1}(\delta_{10}) = \delta_{10}^{-1} +(\delta_{10}^2 + q^2 )^{-1}
-(2\delta_{10}^2 + \frac{1}{2}q^2 )^{-1},~G_n (\delta_{1n}) =\delta_{1n}^2 +q^2,~~n=1,2,\ldots ,
$$
where the corrections $\delta_{1n}$ can be calculated from eqs. (\ref{E:discr}) (\ref{E:ground})
(\ref{E:exc}). In the logarithmic approximation $q\ln q^{-1} \ll 1$, $G_0(\delta_{10}) =2\delta_{10}^2,~~
G_n(\delta_{1n}) =\delta_{1n}^2,~n=1,2,\ldots .$ The quantum number $k_0$ can be found from equation

$$
E^2= \varepsilon_0^2 +p^2k_0^2=\varepsilon_1^2\left [1-\frac{q^2}{(n+\delta_{1n})^2}\right],
$$
with (\ref{E:phi}) for $\varepsilon_{0,1}$.

In conclusion of this section note that the equation absolutely identical to
eqs. (\ref{E:compl}) - (\ref{E:shift}) can be derived by matching the real iteration
functions $v_{1 \mbox {in}}^{(1,2)}(y)$ (\ref{E:iter}) and complex functions of the continuous
states $v_{0+}^{(1,2)}(t)$ (\ref{E:outercont}) having the asymptotic form of the outgoing wave.

\section{Three-subband approximation}\label{S:Three}

In this section we consider the coupling between the discrete states adjacent to the
highest size-quantized level $\varepsilon_{-1}$ and and the continuous states attributed to the
low-lying levels $\varepsilon_{0}$ and $\varepsilon_{1}$. Below we neglect in the set (\ref{E:set1})
the off-diagonal potentials $V_{01}$ and $V_{10}$ describing the interactions of the
$N=0,1$ subbands. Extending the iteration procedure employed above for the single- and double-
subband approximation to the present stage with the trial functions $v_{N}^{(1,2)}=a_{N1,2}$
we arrive at two particular linear independent sixfold vectors
$\vec{V}_{+,-}(v_0^{(1)},v_0^{(2)},v_1^{(1)},v_1^{(2)},v_{-1}^{(1)},v_{-1}^{(2)})$ calculated
for $a_{N+,-}^{(2)}=\pm {\rm i} a_{N+,-}^{(1)},~N=0,1,-1$.
Taking $a_{N+,-}^{(2)}=\\R_N\exp\left[\pm {\rm i} (\zeta_N -\frac{\pi}{2}) \right]$,
we obtain the components $v_N^{(1,2)}$ of the total iteration sixfold vector
$\vec{V}_{\mbox{in}}=\vec{V}_{+}+\vec{V}_{-}$

\begin{eqnarray}\label{E:iter3}
\left.
\begin{array}{l}
v_{0 \mbox{in}}^{(1)}(y)=R_0\sin (Q + \zeta_0) + R_{-1} q \gamma_{0,-1}\cos\zeta_{-1}~;\\
v_{1 \mbox{in}}^{(1)}(y)=R_1\sin (Q + \zeta_1) + R_1 q \gamma_{1,-1}\cos\zeta_{-1}~;\\
v_{-1 \mbox{in}}^{(1)}(y)=R_{-1}\sin (Q + \zeta_{-1}) + R_0 q \gamma_{0,-1}\cos\zeta_{0}
+R_1 q \gamma_{1,-1}\cos\zeta_{1}~,
\end{array}
\right \}
\end{eqnarray}
where $Q(y)$ is determined in eq. (\ref{E:iter}) and $R_N$ and $\zeta_N$ are
arbitrary constants and phases, respectively. The parameter
$\gamma_{0,-1}=\gamma_{0,1}$ (\ref{E:couple}), while

\begin{eqnarray}\label{E:couple1}
2\pi \gamma_{1,-1}=\cos 2\alpha_0\left[-\mbox{Si}(\pi + 2\alpha_0)- \mbox{Si}(\pi - 2\alpha_0) \right]
\nonumber\\
+\sin 2\alpha_0 \left[\mbox{Ci}(\pi + 2\alpha_0)- \mbox{Ci}(\pi - 2\alpha_0) \right],~\alpha_0=\frac{\pi x_0}{d}
\end{eqnarray}
describes the coupling induced by the potentials $V_{-11}=V_{1-1}$ (\ref{E:pot}).
The functions $v_{N\mbox{it}}^{(2)}(y)$ can be obtained from the functions
 $v_{N\mbox{it}}^{(1)}(y)$ (\ref{E:iter3}), respectively, by replacing
$\sin (Q + \zeta_N)~\mbox{by}~\cos (Q + \zeta_N), ~\cos (\zeta_N)~\mbox{by}~\sin (\zeta_N)
~\mbox{and}~q~\mbox{by}~-q$.

As mentioned above further we match the wave functions of the continuous spectrum $v_N^{(1,2)}(y),~N=0,1$
having the asymptotic form of the outgoing wave $v_{N+}^{(1,2)}$ (\ref{E:outercont}) to give in turn
the sixfold vector $\vec{V}_{\mbox{out}}$ with the components

\begin{eqnarray}\label{E:outercont3}
v_{N\mbox{out}}^{(1)}(y)=(1+c_N)\exp\left[{\rm i} \left(q\ln2k_N y +\xi_{+}
 \right)\right]
 \nonumber\\
  -(1-c_N)\exp\left[-\imath \left(q\ln2k_N y +\xi_{-}
 \right)  \right],~N=0,1,
\end{eqnarray}
where $c_N$ and $\xi_{+,-}$ are given in eqs. (\ref{E:outercont2}) (\ref{E:outercont1}) and (\ref{E:outercont}).
The wave functions $v_{N\mbox{out}}^{(2)}(y),~N=0,1$ can be obtained from
the functions $v_{N\mbox{out}}^{(1)}(y)$ (\ref{E:outercont3}), respectively, by replacing
$\xi_{+,-}~\mbox{by}~\xi_{+,-}+\frac{\pi}{2}.$ The wave functions
$v_{-1\mbox{out}}^{(1,2)}(y)$ have the form (\ref{E:outer1}), in which $\nu_{N}$ (\ref{E:outer})
and $\Theta_N$ (\ref{E:theta}) are calculated for $N=-1$.

Matching the sixfold wave vectors $\vec {V}_{\mbox{in}}~\mbox{and}~\vec {V}_{\mbox{out}}$ within
the intermediate region by imposing the conditions

$$
\frac{v_{N\mbox{out}}^{(1)}}{v_{N\mbox{out}}^{(2)}}=\frac{v_{N\mbox{in}}^{(1)}}{v_{N\mbox{in}}^{(2)}},
\qquad N=0,1,-1,
$$
where $v_{N\mbox{out}}^{(1,2)}$ are given by eqs. (\ref{E:outercont3}) and (\ref{E:outer}) and
$v_{N\mbox{in}}^{(1,2)}$ by eqs. (\ref{E:iter3}) we obtain

\begin{eqnarray}\label{E:setcouple1}
\left.
\begin{array}{l}
R_0 [A_0^{(-)} \cos(Q + \zeta_0) - {\rm i} A_0^{(+)}\sin(Q + \zeta_0)]
 - R_{-1} q \gamma_{0,-1}[ A_0^{(-)}\sin\zeta_{-1}+{\rm i} A_0^{(+)}\cos\zeta_{-1}]=0~;\\
R_1 [A_1^{(-)} \cos(Q + \zeta_1) - {\rm i} A_1^{(+)}\sin(Q + \zeta_1)]
 - R_{-1} q \gamma_{1,-1}[ A_1^{(-)}\sin\zeta_{-1}+{\rm i} A_1^{(+)}\cos\zeta_{-1}]=0~;\\
R_{-1}\sin(\omega_{-1}-Q - \zeta_{-1})-R_{1}q\gamma_{1,-1}\cos(\omega_{-1}-Q - \zeta_{1})-
R_{0}q\gamma_{0,-1}\cos(\omega_{-1}-Q - \zeta_{0})=0~,
\end{array}
\right \}
\end{eqnarray}
In eqs. (\ref{E:setcouple1})

$$
A_N^{(+,-)}=(1+c_{N})\exp\left[{\rm i} \left(q\ln2k_N y +\xi_{+}
 \right)\right]
 \pm(1-c_{N})\exp\left[-{\rm i} \left(q\ln2k_N y +\xi_{-}
 \right)\right]
$$
and $Q(y)~\mbox{and}~\omega_{-1}(y)$ are introduced by eqs. (\ref{E:iter}) and (\ref{E:outer1}) for
$N=-1$, respectively.

Solving the set (\ref{E:setcouple1}) by the determinantal method we obtain the equations for the
complex quantum numbers $\eta_N=2qE/p\nu_{-1}$ (\ref{E:outer}), in which we take for the phases
$\zeta_N = \frac{\pi}{2},~N=0,1,-1$ and keep the terms of the first order of $q\ll 1$

\begin{equation}\label{E:image1}
\omega_{-1} - Q - \frac{\pi}{2} = q^2\sum_{N=0,1}\gamma_{N,-1}\left( qA_{N,-1} +\rm {i} B_{N,-1}\right),
\end{equation}
with

$$
A_{N,-1} = B_{N,-1}^2\left (\ln\frac{k_N d}{2} + \ln\frac{D}{d} +C -1 \right);
~~B_{N,-1}=\frac{2\varepsilon_N}{pk_N}.
$$
and the quantum numbers $k_{0,1}$ are obtained from

$$
E^2= \varepsilon_0^2 +p^2k_0^2=\varepsilon_1^2 +p^2k_1^2 =\varepsilon_{-1}^2\left [1-\frac{q^2}{(n+\delta_{-1n})^2}\right]
$$
with (\ref{E:phi}) for $\varepsilon_{-1}$.

The complex quantum numbers $\eta$ calculated from eq. (\ref{E:image1}) lead to the complex impurity
energy levels adjacent to the size-quantized second excited sub-band $\varepsilon_{-1}$

\begin{equation}\label{E:compl1}
E_{-1n}=\varepsilon_{-1} - \varepsilon_{-1}\frac{q^2}{2(n+\delta_{-1n})^2}
 +W_{-1n} - \imath \frac{\Gamma_{-1n}}{2},~~n=0,1,2,\ldots
\end{equation}
where the resonant width $\Gamma_{-1n}$ and shift $W_{-1n}$ have the form

\begin{equation}\label{E:widthn}
\Gamma_{-1n} = 2\varepsilon_{-1}\frac{q^2}{(n+\delta_{-1n})^3}G_n
q\left(\gamma_{1,-1}^2 B_{1,-1} + \gamma_{0,-1}^2 B_{0,-1} \right)
\end{equation}
and

\begin{equation}\label{E:shiftn}
W_{-1n} =- \varepsilon_{-1}\frac{q^2}{(n+\delta_{-1n})^3}G_n
q^2\left(\gamma_{1,-1}^2 A_{1,-1} + \gamma_{0,-1}^2 A_{0,-1} \right).
\end{equation}
The coefficients $G_n$ are defined in eqs. (\ref{E:width}) and (\ref{E:shift}).

\section{Discussion}\label{S:Disc}

We define the binding energy of the electron $E_{Nn}^{(b)}$ in the $n$-th quasi-Coulomb
state associated with the $N$ size-quantized subband as the real part of the difference
between the size-quantized energy $\varepsilon_N$ (\ref{E:phi}) of the free electron and
the energy of the impurity electron $E_{Nn}$ given by eqs. (\ref{E:rydberg}), (\ref{E:compl})
and (\ref{E:compl})
for the ground $N=0$, first $N=1$ and second $N=-1$ excited subbands, respectively.
Since the resonant shifts
$W_{Nn}$ (\ref{E:shift}), (\ref{E:shiftn})
are of the order of $q^2 \ll 1~(W_{0n}=0)$ with respect
to the Rydberg energies determined by the second terms in the right-hand parts of eqs.
(\ref{E:rydberg}), (\ref{E:compl}) and (\ref{E:compl}) the binding energies read

\begin{eqnarray}\label{E:bind}
E_{Nn}=
\left\{
\begin{array}{ll}
\varepsilon_N \frac{q^2}{2(n + \delta_{Nn})^2};\, &n=1,2,\ldots \\
\varepsilon_N\left[1 - \frac{1}{{\sqrt{1+\frac{q^2}{\delta_{N0}^2}}}}\right];~&n=0
\end{array}
\right.
\end{eqnarray}
where the corrections $\delta_{Nn}$ can be calculated in the single-subband approximation
from eqs. (\ref{E:ground}) and (\ref{E:exc}) for the ground $n=0$ and excited $n=1,2,\ldots$
impurity states, respectively. It follows from eqs. (\ref{E:bind}) and (\ref{E:phi}) that
the binding energy $E_{Nn}^{(b)}\sim \varepsilon_N \sim d^{-1}$
and the oscillatory part of $\varepsilon_N$ (see $\tilde{\sigma}(d)$ in eq. (\ref{E:phi})) decrease
with increasing the ribbon width $d$.  In an effort to render our calculations
close to an experimental setup, we take below for the estimates of the expected values the parameters
$q=0.13~(\epsilon \simeq 25)$ \cite{rob} and $q=0.24~(\epsilon \simeq 10)$ corresponding to
the $\mbox{HfO}_2$ and sapphire, respectively, employed as substrates for GNR \cite{han}.
The latter parameter $q$ is close to the limit caused by the condition $z_0 \ll 1$
(see below eq.(\ref{E:rydberg})). Further we focus on the monotonic dependence
$\sim d^{-1}$ and keep the the parameter $\tilde{\sigma}$ in eq. (\ref{E:phi})
for the levels $\varepsilon_N$ to be $\tilde{\sigma}\simeq 0.3$.

\begin{figure}[t!]
\begin{center}
\includegraphics*[width=.7\columnwidth]{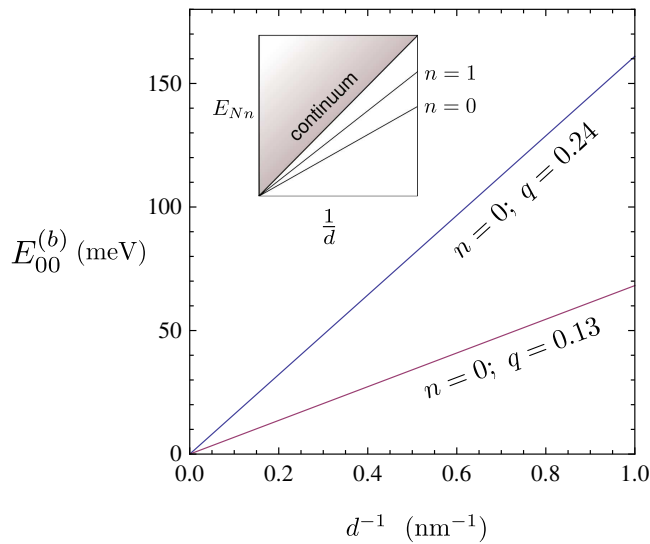}
\end{center}
\caption{\label{fig1} The binding energy $E_{0n}^{(b)}$ (\ref{E:bind}) of the ground state
$n=0$ calculated for($q=0.13;~0.24$)
as a function of the reciprocal width $\frac{1}{d}$ of the GNR. Impurity is placed symmetrically
to the boundaries $(x_0 =0)$. The parameter $\tilde{\sigma}=0.3$.}
\end{figure}

The dependencies of the binding energies on the width of the GNR $d$ for the ground
state for different strengths of the
impurity potential are given in Fig. 2. These graphs,
while ignoring the oscillations,
are qualitatively completely in line
with the data of the numerical calculations and experimental observations
recently performed with the related Coulomb systems. The exciton effects
in the armchair GNRs were studied in frame of the tight-binding model
\cite{jia} and density functional theory \cite{zhu}, while Han et.al. \cite{han}
investigated experimentally the influence of the localized states in GNRs
on the electron transport. The relation $E^{(b)}\sim d^{-1}$ including oscillations
\cite{jia, zhu} have been found to occur. The differences between
the impurity states considered here and the exciton and localized
states prevent us from a detailed quantitative comparison.

\begin{figure}[t!]
\begin{center}
\includegraphics*[width=.7\columnwidth]{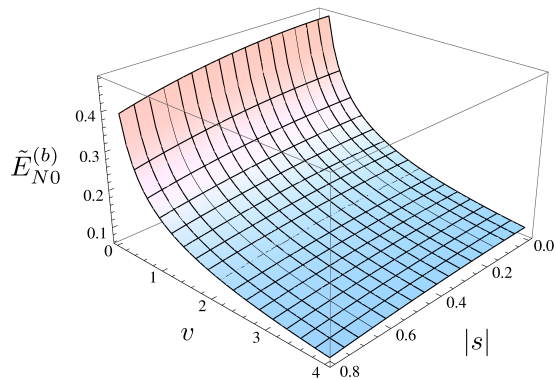}
\end{center}
\caption{\label{fig1} The dimensionless binding energy $\tilde{E}_{N0}^{(b)}=E_{N0}^{(b)}/\pi p d^{-1}$
calculated from (\ref{E:bind}), (\ref{E:ground}), (\ref{E:phi}) with $q=0.20$
for the ground state $(n=0)$
plotted as a function of the effective index of the corresponding subband $v=|N-\tilde{\sigma}|$
and the relative impurity position $s=\frac{2x_0}{d}$ in the GNR of width $d$.}
\end{figure}

The Coulomb pattern of the energy levels (\ref{E:rydberg}) enables
to introduce the effective Rydberg constant $Ry_N$, the Bohr radius $a_{0N}$
and the mass $M_N$ for the impurity electron in GNR

$$
Ry_N = \frac{q^2 |N-\tilde{\sigma}|\pi p}{2d},~a_{0N}=\frac{d}{\pi|N-\tilde{\sigma}|q},~
M_N = \frac{|N-\tilde{\sigma}|\pi \hbar^2}{pd}.
$$
which additionally illustrate the physical reason of the bonding
of the impurity electron, namely the quasi-1D geometry of the GNR.
Note that the bound states arise at any finite width
$d<\infty$. This result is qualitatively analogous to the effect of anti-diamagnetism
caused by the influence of the magnetic field on the weakly bound atomic state. Demkov and
Drukarev \cite{demdruk} considered the 3D potential well of small radius and depth
to provide the capturing of the electron. It was shown that the arbitrarily weak
magnetic field
$B$ induces the bound electron state with the binding energy $E^{(b)}\sim a_{B}^{-2}~
(a_{B}=(\hbar /eB)^{1/2}~\mbox{is the magnetic length})$. The common reason for this is that
the finite width $d<\infty$ and magnetic length $a_B < \infty$ transform the graphene
monolayer and atomic structure, respectively, into the quasi-1D systems, which are more favorable
to generate bound states. The dependencies $E^{(b)}\sim a_{B}^{-2}$
and $E^{(b)}\sim d^{-1}$ correspond to the different dispersion laws namely
$E^{(b)}\sim p^2$ and $E^{(b)}\sim p$ $(p\simeq \hbar /r~\mbox{is the momentum})$ for the atomic
($r\simeq a_{B}$) and GNR ($r\simeq d$) electron, respectively.

The dependence of the binding energy $E_{Nn}^{(b)}$ (\ref{E:bind}) on the displacement
of the impurity centre $x_0$ from the mid-point of the ribbon $x=0$ is contained in the
corrections $\delta_{Nn}(x_0)$ namely in the term $\ln Dd^{-1}$ in eqs.
(\ref{E:ground}) and (\ref{E:exc}), while $E_{Nn}^{(b)}$ as a function of the
effective number of the subband $|N-\tilde{\sigma}|$ is given by the sub-band threshold
$\varepsilon_N$ (\ref{E:phi}) mainly and the term $\ln |N-\tilde{\sigma}|$ in the
correction $\delta_{Nn}$.
The dimensionless binding energy
$E_{N0}^{(b)}/\pi p d^{-1}$ as a function of the effective quantum number $|N-\tilde{\sigma}|$
and relative displacement $x_0 / (d/2)$ for the ground $n=0$ state
is depicted in Fig. 3. Clearly,
the higher the subband i.e. the
greater the value $|N-\tilde{\sigma}|$ is the less the binding energy $E_{N0}^{(b)}$.
Also
the binding energy decreases when the impurity shifts from the mid-point of the ribbon towards
the boundaries. The latter conclusion coincides with those obtained for the quantum well
in Refs.
\cite{bast,blom,mailh,tanaka,greene}.

The inter-band coupling shifts the strictly discrete excited Rydberg
series $E_{Nn}~(N\neq 0)$ (\ref{E:rydberg}) calculated in single-subband approximation
by an amount $W_{Nn}$ (\ref{E:shift}) $N=1$, and (\ref{E:shiftn}) $N=-1$
and transforms them to the
quasi-discrete levels of width $\Gamma_{Nn}$ (\ref{E:width}) $N=1$,
and (\ref{E:widthn}) $N=-1$. Note that the conclusions made on the base of the
first and second excited subbands can be qualitatively extended to others.
Since the resonant shifts $W_{Nn}\sim q^2$ first are much less than the resonant widths
$\Gamma_{Nn}\sim q$ at $q\ll 1~(W_{Nn} \ll \Gamma_{Nn})$ and second the resonant shifts do not change
the discrete character of the energy spectrum (\ref{E:rydberg}) we focus on the widths $\Gamma_{Nn}$.
It is clear from eqs. (\ref{E:width}), and (\ref{E:widthn}) that the widths
$\Gamma_{Nn}\sim\varepsilon_N \sim d^{-1}$ increase with decreasing the ribbons width $d$.
Note that this dependence is opposite to that in a semiconductor narrow quantum well:
the narrower the well is the less are the resonant widths \cite{yen,blom,monschm05}.
The reason for this is that in the quantum well the resonant width
$\Gamma_{Nn}\sim E_{Nn}^{(b)} \left( E_{Nn}^{(b)}/\Delta \varepsilon_N  \right)^2$
where the impurity Rydberg constant $Ry \simeq E_{Nn}^{(b)}$ and the binding energy $E_{Nn}^{(b)}$
do not depend on the well width $d$, while the inte-rband energy distance
$\Delta \varepsilon_N \sim d^{-2}$ increases and consequently
the resonant width decreases with the narrowing of the quantum well. For the ribbon
$E_{Nn}^{(b)} \sim \varepsilon_N \sim \Delta \varepsilon_N \sim d^{-1}$ (\ref{E:bind})
and the inter-subband coupling do not depend on the ribbon width $d$ and
$\Gamma_{Nn}\sim E_{Nn}^{(b)}\sim \varepsilon_N \sim d^{-1}$.

\begin{figure}[t!]
\begin{center}
\includegraphics*[width=.7\columnwidth]{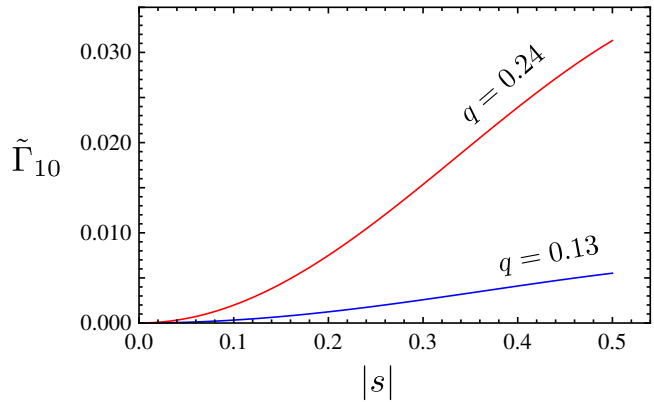}
\end{center}
\caption{\label{fig1} The resonant width $\Gamma_{10}$ (\ref{E:width}) of the ground impurity state $(n=0)$
relatively to the corresponding threshold $\varepsilon_1$ (\ref{E:phi})
$(\tilde{\Gamma}_{10}=\Gamma_{10}/\varepsilon_1)$ versus the relative
impurity position $s=\frac{2x_0}{d}$ in the GNR of width $d$
providing the parameter $\tilde{\sigma}=0.3$. The parameter $q$ is taken to be
$q=0.13;~0.24$.}
\end{figure}

The dependence of the resonant widths $\Gamma_{1n}(x_0)$ (\ref{E:width}),
calculated in the double-subband approximation, on the position of the impurity centre $x_0$
 is described by the coupling parameter
$\gamma_{01}$ (\ref{E:couple}) and the corrections $\delta_{1n}$
(\ref{E:ground}) and (\ref{E:exc}).
The dependencies of the relative resonant widths $\Gamma_{1n}/\varepsilon_1$
(\ref{E:width}), on the
dimensionless shift $s=2 x_0 /d$ for the ground $n=0$ state
are presented in Fig.4, in which the limitation on the parameter
$s$ are caused by
the condition imposed on $z_0$ placed below eq. (\ref{E:rydberg}).
For the impurity positioned at the mid-point of the ribbon
$x_0 =0$ the resonant width and shift both vanish $(\Gamma_{1n}(0)=W_{1n}(0)=0)$ because
of the even $x$-parity of the Coulomb potential $V(\vec{\rho})$ (\ref{E:coulomb})
in eq. (\ref{E:pot}) and opposite parities of the neighboring $N=0,1$
transverse $x$-states to give
$V_{01}=\gamma_{0,1}=0$. Both in the quantum well and in the ribbon the shift
of the impurities from their
mid-points eliminates the even $x$-parity of the potential
$V(\vec{\rho})$ (\ref{E:coulomb}) in eq. (\ref{E:pot}),
that leads to the coupling
between the $N=0~\mbox{and}~N=1$ subbands.
If the impurity displaces from the mid-point towards
the boundaries $|x_0|=d/2$ the resonant widths $\Gamma_{1n}$ (\ref{E:width})
monotonically increases. This correlates completely with the analogous dependence
found for the impurity states in the semiconductor quantum well \cite{yen,blom,monschm05}.
For small shifts $\alpha_0 \ll 1$ in eq. (\ref{E:couple}) we obtain for the
parameter $\gamma_{0,1}$ in eqs. (\ref{E:width})
$\gamma_{0,1}=(\frac{2}{\pi})Si(\frac{\pi}{2}) \alpha_0\ll 1$
while for the impurity positioned close to the ribbon edge $x_0 \simeq \frac{d}{2}$
we obtain $\gamma_{0,1}=\frac{1}{\pi}Si(\pi)~\mbox{with}~Si(\frac{\pi}{2})= 1.37,~Si(\pi)=1.85$.
Note that the zeroth width of the first excited $(N=1)$
$n-$ series in case of the symmetrical
$x_0=0$ impurity position is a consequence of the double-subband approximation. In the multi-
subband approximation the levels of the above mentioned series would acquire finite widths.

\begin{figure}[t!]
\begin{center}
\includegraphics*[width=.7\columnwidth]{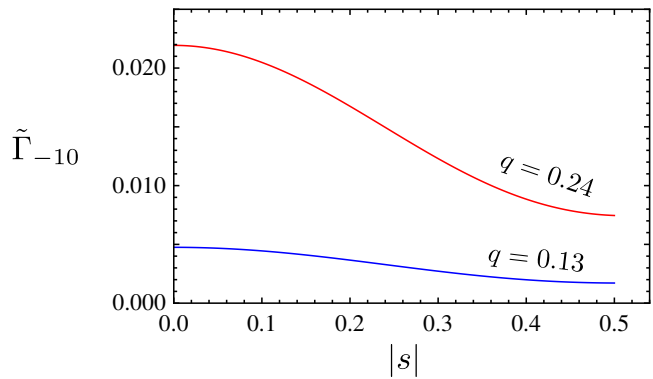}
\end{center}
\caption{\label{fig1} The resonant width $\Gamma_{-10}$ (\ref{E:widthn}) of the ground impurity state $(n=0)$
relatively to the corresponding threshold $\varepsilon_{-1}$ (\ref{E:phi})
$(\tilde{\Gamma}_{-10}=\Gamma_{-10}/\varepsilon_{-1})$ versus the relative
impurity position $\frac{2x_0}{d}$ in the GNR of width $d$
using the parameter value $\tilde{\sigma}=0.3$. The parameter $q$ is taken to be
$q=0.13;0.24$.}
\end{figure}

The dependence of the resonant widths (\ref{E:widthn}) of the impurity states corresponding
 to the second excited subband $N=-1$ on the position of the impurity centre
 is completely different from that related to the first excited subband $N=1$.
 Equation (\ref{E:widthn}) shows that contributions to the resonant widths $\Gamma_{-1n}$
 are caused by the coupling with the subbands $N=0~(\sim \gamma_{0,-1}^2 )$ (\ref{E:couple})
 and $N=1~(\sim \gamma_{1,-1}^2 )$ (\ref{E:couple1}). Note that the estimated
 contribution to the resonant width $\Gamma_{-1n}$ caused by the neglected
 coupling between the $N=0~\mbox{and}~N=1$ subbands is of the order of $q\gamma_{01}^2 \ll 1.$
 In the vicinity of the mid-point
 $(|x_0|\ll d/2,~ \alpha_0 \ll 1)$ the subband $N=1$ contributes mostly $(\gamma_{0,-1}\simeq 0,~
 \gamma_{1,-1}\simeq - \frac{1}{\pi}Si(\pi))$, while for
 $|x_0| \simeq d/2,~\alpha_0 \simeq \pi/2$ both subbands
 contribute
 $\gamma_{0,-1}\simeq \frac{1}{\pi}Si(\pi),~\gamma_{1,-1}\simeq \frac{1}{2\pi}Si(2\pi)$.
 The position of the impurity $\bar{x}_0=\frac{d}{\pi}\bar{\alpha}_0$,
 at which the effects of the subbands $N=0$ and
$N=1$ on the resonant width $\Gamma_{-1n}$ are in balance is determined by the root $\bar{\alpha}_0$
of the equation

$$
\gamma_{1,-1}^2B_{1-1}=\gamma_{0,-1}^2B_{0-1}.
$$
to give the result $\bar{\alpha}_0 =0.58,~ \bar{x}_0 =0.37\frac{d}{2}$. The coupling between the subbands
$N=-1~\mbox{and}~N=1$ provides the nonzero widths $\Gamma_{-1n}$ and shifts $W_{-1n}$
for any positions $x_0$ of the impurity. The width $\Gamma_{-10}$ as a function of the impurity
shift $x_0$ is given in Fig.5 demonstrating the monotonic drop within the same regions
as those corresponding to Fig.4.

As mentioned above the presented method is valid under the conditions $q\ll 1$
for the excited impurity states $n=1,2,\ldots$ and $z_0 (q)\ll 1$ (\ref{E:ground})
for the ground state $n=0$. Under these conditions the radius
of the impurity state considerably exceeds the width of the GNR
so that the ribbon is narrow compared to the impurity size.
It follows from eq. (\ref{E:discr}) that a
"big logarithm" can only hardly achieved \cite{perpop,pop70,pop71,zeld}
i.e. the logarithmic approximation $|\ln q|\gg 1$
ensures the real smallness of $z_0$. However the previous
calculations related to the ground state of the quasi-1D diamagnetic exciton
\cite{zhilkyun} and present estimates show that a reasonably small parameter
 $q$ leads to values $z_0 < 1$, which provide a quite accurate and adequate
 description of the ground impurity state in GNR.

 Taking into account possible experiments we estimate the expected electron
 binding energy for the impurity centre placed at the middle point of the GNR
 of width 1 nm on the sapphire substrate as
 $E_{00}^{(b)}\simeq 160~\mbox{meV}$ and on the $\mbox{HfO}_2$ substrate
 as $E_{00}^{(b)}\simeq 68~\mbox{meV}$. This is less than the data attributed
 to the $\mbox{SiO}_2$ substrate $(\epsilon =3.9)$ because of the relatively small screening
 of the impurity potential. Also an estimate of the lifetimes $\tau_{Nn}=\hbar/\Gamma_{Nn}$
 yields for the impurity positioned at the mid-point of the GNR
 $\tau_{-10}\simeq0.21 ~\mbox{ps}~\mbox{and}~0.049~\mbox{ps}$ for the $~\mbox{HfO}_2$ and sapphire
 substrate, respectively. For the $~\mbox{SiO}_2$ substrate
  the screening of the impurity attraction is less,
 the lifetime is reduced and therefore less favourable for
 a corresponding experimental observation.
A shift of the impurity centre $|x_0| \simeq 0.4 d/2 $ generates lifetimes $\tau_{10}$ of the
same order as $\tau_{-10}$ at $x_0=0$. The electrons captured onto such short-lived trap states
 will most likely contribute to the dc transport. However, the high-frequency
 response of such electrons may reveal the signatures of localization.

Clearly, the above considered quasi-Rydberg series (\ref{E:rydberg}) do not cover the total
set of discrete states. The oscillations of the wave functions (\ref{E:iter}) and (\ref{E:outer1})
caused by the logarithmic term are an indicator of additional energy levels positioned below the series
(\ref{E:rydberg}). Since the possible strong shift of these levels away from the threshold
$\varepsilon_N$ is against the spirit of the employed adiabatic perturbation theory
$(q\ll 1)$ describing the shallow energy levels we are limited to qualitative estimates
based on the quasi-classical relativistic approach \cite{Shyt1} and \cite{per1}.

\begin{figure}[t!]
\begin{center}
\includegraphics*[width=.7\columnwidth]{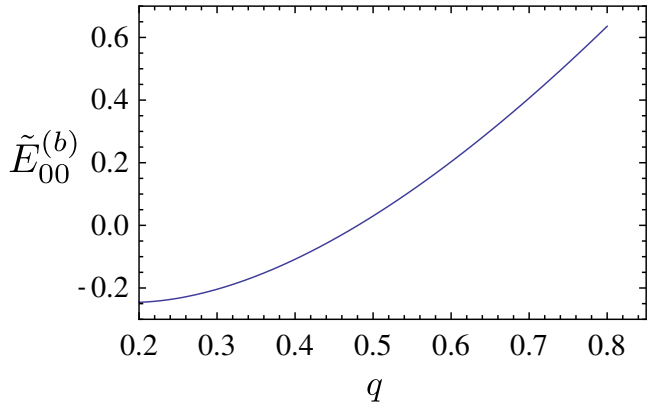}
\end{center}
\caption{\label{fig1} The dimensionless
binding energy $\tilde{E}_{00}^{(b)}=E_{00}^{(b)}/\varepsilon_0$
of the quasi-classical ground state $(N=n=0)$ found from
(\ref{E:class}) for $E_{00}$ and from (\ref{E:phi})
with $\tilde{\sigma}=0.3$ for $\varepsilon_0$
versus the parameter $q$.}
\end{figure}

In an effort to elucidate the origin of this additional series
let us consider the so called "logarithmic" energy levels governed by the logarithmic
potential (\ref{E:limit}) taken for $d_1 =d_2=d~(x_0 =0)$.
These levels can be calculated from the Bohr-Sommerfeld quantization rule

\begin{equation}\label{E:bohrsomm}
\int_0^{y_0}\mathcal{P}(y)dy=\pi\hbar(n+\frac{1}{2});~n=0,1,2,\ldots,
\end{equation}
where $\mathcal{P}(y)$ and $y_0 \ll d$ are the classical relativistic momentum
and turning point, respectively, with

\begin{eqnarray}
\mathcal{P}^2(y) &=&
\frac{1}{v_F^2} \left[\left( E - 2\frac{\beta}{d}\ln\frac{y}{d}\right)^2 - \varepsilon_N^2 \right];
\nonumber\\
\mathcal{P}(y_0)&=&0.
\end{eqnarray}

Equation (\ref{E:bohrsomm}) admits an exact solution which provides for the energies

\begin{equation}\label{E:class}
E_{Nn}=\frac{2p}{d}q\left[ \ln\frac{(n+\frac{1}{2})}{|N-\tilde{\sigma}|}
-\ln K_1(s_N)\right],~s_N=\frac{\varepsilon_N d}{2pq}=\frac{|N-\tilde{\sigma}|\pi}{2q},
\end{equation}
where $K_1(s)$ is the modified Bessel function \cite{abram}.

The binding energy $E_{Nn}^{(b)}=\varepsilon_N - E_{Nn}$ with $E_{Nn}$ calculated from
(\ref{E:class}) reads $E_{Nn}^{(b)} \sim q\ln q$ both for $q\ll 1~(s_N \gg 1)$ and for
$q\gg 1~(s_N \ll 1,~|\ln q| \gg 1)$. It follows that the weakness of the logarithmic
singularity and smallness of the strength of the impurity potential
$(q\ll 1)$ seems not to provide the bonding of the quasi-classical
relativistic electron $(E_{Nn}^{(b)}< 0)$, while a sufficiently strong attraction $(q \leq 1)$
could produce a localized impurity state $(E_{Nn}^{(b)}>0)$. The dependence
of the binding energy of the quasi-classical ground state $(N=n=0)$ found from
(\ref{E:class}) on the parameter $q$ is depicted in Fig.6. The ground
"logarithmic" level arises at the critical value $q_0 \simeq 0.48$ and shifts
towards lower energies to provide for the binding energy
$0.1 < E_{00}^{(b)}/\varepsilon_0<0.5~\mbox{for}~0.54 < q < 0.74$. The above
can be considered as no more than only a qualitative evidence of existence
of such additional states in GNR that have transformed from the
collapsed states in the graphene
monolayer governed by the 2D impurity potential $\sim - r^{-1}$ \cite{Shyt1}.
Though the "logarithmic" and quasi-Rydberg levels in principle correspond to the same region
of the parameter $q < 1$ any numerical comparison between the
results for the quasi-Rydberg series based on the Dirac equation
and those for the "logarithmic" levels derived from the quasi-classical
method applied moreover to the ground state seems to be incorrect. The total
set of the impurity states in GNR requires a further study of the equations
(\ref{E:set}) with the potential (\ref{E:pot}), having the logarithmic singularity in
the vicinity of the impurity centre.

\section{Conclusion}\label{S:Concl}

We have developed an analytical adiabatic approach to the problem of
bound and meta-stable (Fano resonances) quasi-Coulomb impurity states in a
narrow gaped armchair graphene nanoribbon (GNR). The width of the GNR is taken
to be much less than the radius of the impurity. This adiabatic criterion implies
a variable width of the GNR and simultaneously the smallness
of the Coulomb interaction relative to the size-quantized energy
induced by the GNR.
The energy spectrum of the impurity electron is a sequence of the
series of the quasi-Rydberg discrete and resonant states adjacent to the
ground and excited size-quantized subbands, respectively. The binding energies
and the resonant widths and shifts attributed to the
inter-subband coupling are calculated in an explicit form in the single-
and multi-subband approximation, respectively. The binding energies
and the resonant widths both increase with decreasing the GNR width. As the impurity
centre displaces from the mid-point of the GNR the binding energies
decrease, while the resonant widths of the quasi-Rydberg series associated
with the first/second excited sub-bands increase/decrease, respectively.
Our analytical
results are in complete agreement with those found by other theoretical
approaches and in particular numerical studies. Estimates of the expected values
show that the bound and meta-stable impurity states in GNR can be observed
experimentally.

\section{Acknowledgments}\label{S:Ackn}
The authors are grateful to C. Morfonios for technical assistance.
Financial support by the Deutsche Forschungsgemeinschaft is gratefully acknowledge.


\begin{thebibliography}{99}


\bibitem{Castro}
A.~H.~Castro Neto, F.~Guinea, N.~M.~R.~Peres, K.~S.~Novoselov, and A.~K.~Geim,
Rev. Mod. Phys. \textbf{81} 109 (2009)


\bibitem{Rosl}
O.~Roslyak, G.~Gumbs, and D.~Huang. Phil. Trans. R. Soc. A \textbf{368}, 5431 (2010)


\bibitem{Namura}
K.~Namura and A.~H.~MacDonald, Phys. Rev. Lett. \textbf{98}, 076602 (2007)


\bibitem{Peres}
N.~M.~R.~Peres, F.~Guinea and A.~H.~Castro Neto, Phys. Rev. B \textbf{73}, 125411 (2006)


\bibitem{Bis}
R.~B.~Biswas, S.~Sachdev, and D.~T.~Son,
Phys. Rev. B \textbf{76}, 205122046803 (2007)


\bibitem{Nov}
D.~S.~Novikov, Phys. Rev. B \textbf{76}, 245435 (2007)


\bibitem{Per}
V.~M.~Pereira, J.~Nilsson, and A.~H.~Castro Neto, Phys. Rev. Lett. \textbf{99}, 166802 (2007)


\bibitem{Shyt}
A.~V.~Shytov, M.~I.~Katsnelson, and L.~S.~Levitov, Phys. Rev. Lett. \textbf{99}, 236801 (2007)


\bibitem{Shyt1}
A.~V.~Shytov, M.~I.~Katsnelson, and L.~S.~Levitov, Phys. Rev. Lett. \textbf{99}, 246802 (2007)


\bibitem{landau}
L.~D.~Landau, and E.~M.~Lifshitz, \emph{Quantum Mechanics: Non-Relativistic Theory}
(Pergamon, London) 1981


\bibitem{perpop}
A.~M.~Perelomov and V.~S.~Popov, Theor.Math.Phys. \textbf{4}, 664 (1970)


\bibitem{pop70}
V.~S.~Popov, Phys.Atom.Nucl. Phys. \textbf{12}, 429 (1970)


\bibitem{pop71}
V.~S.~Popov, JETP \textbf{60}, 1228 (1971)


\bibitem{zeld}
Ya.~B.~Zeldovich and V.~S.~Popov, Sov.Phys.Usp.\textbf{14}, 673 (1972)


\bibitem{gupta}
K.~S.~Gupta, S~Sen, Phys. Rev. B \textbf{78}, 205429 (2008)


\bibitem{gupta1}
K.~S.~Gupta, S~Sen, Mod. Phys. Lett. A \textbf{24}, 99 (2009)


\bibitem{harr}
P.~Harrison, \emph{Quantum Wells, Wires and Dots} (Wiley New York, 2000)


\bibitem{monschm09}
B.~S.~Monozon, P~Schmelcher, Phys. Rev. B \textbf{79}, 165314 (2009)


\bibitem{demdruk}
Yu.~N.~Demkov, G.~P.~Drukarev, Sov. Phys. JETP-USSR \textbf{22}, 182 (1966)

\bibitem{keld}
L.~V.~Keldysh, JETP Lett. \textbf{29}, 658 (1978)


\bibitem{katsn}
M.~I.~Katsnelson, K.~S.~Novoselov, and A.~K.~Geim,Nature Phys. \textbf{2}, 620 (2006)


\bibitem{katsnnov}
M.~I.~Katsnelson, and K.~S.~Novoselov, Solid State Commun.\textbf{143}, 3 (2007)


\bibitem{Brey}
L.~Brey and H.~A.~Fertig, Phys. Rev. B \textbf{73}, 235411 (2006)


\bibitem{fano}
U.~Fano, Phys. Rev. \textbf{124}, 1866 (1961)


\bibitem{hwang}
E.~H.~Hwang and S.Das~Sarma, Phys. Rev. B \textbf{75}, 205418 (2007)


\bibitem{hashow}
H.~Hasegawa and R.~E.~Howard, J.~Phys. Chem. Solids \textbf{21}, 173 (1961)


\bibitem{monschm05}
B.~S.~Monozon and P.~Schmelcher, Phys. Rev.~B \textbf{71}, 085302 (2005)


\bibitem{monschm07}
B.~S.~Monozon and P.~Schmelcher, Phys. Rev.~B \textbf{75}, 245207 (2007)


\bibitem{berlif}
V.~B.~Berestetskii, E.~M.~Lifshitz, L.~P.~Pitaevskii, \emph{Quantum Electrodynamics},
Butterworth-Heinemann, Oxford, Second Edition, (1982)


\bibitem{abram}
\emph{Handbook of Mathematical Functions}, edited by M.~Abramowitz and
I.~A.~Stegun (Dover, New York, 1972)


\bibitem{baterd}
\emph{Higher Transcendental Functions v.1}, edited by H.~Bateman and
A.~Erdelyi (Mc Graw-Hill Book Company, Inc., New York, Toronto, London 1953)


\bibitem{newton}
R.~G.~Newton\emph{Scattering Theory of Waves and Particles}
(Springer, New York, 1982)


\bibitem{rob}
J.~Robertson, Eur. Phys. J. Appl. Phys.\textbf{28}, 265 (2004)


\bibitem{han}
M.~Y.~Han, J.~C.~Brant, and P.~Kim, Phys. Rev. Lett. \textbf{104}, 056801 (2010)


\bibitem{jia}
Y.~L.~Jia, X.~Geng, H.~Sun, and Y.~Luo, Eur. Phys. J. B \textbf{83}, 451 (2011)


\bibitem{zhu}
X.~Zhu and H.~Su, J. Phys. Chem. A \textbf{115}, 11998 (2011)


\bibitem{bast}
G.~Bastard, Phys. Rev. B \textbf{24}, 4714 (1981)


\bibitem{blom}
A.~Blom, M.~A.~Odnobludov, I.~N.~Yassievich, and K.~A.~Chao, Phys. Rev. B \textbf{68}, 165338 (2003)


\bibitem{mailh}
C.~Mailhiot, Y.~-C.~Chang, and T.~C.~McGill, Phys. Rev. B \textbf{26}, 4449 (1982)


\bibitem{tanaka}
K.~Tanaka, M.~Nagaoka, and T.~Yamabe, Phys. Rev. B \textbf{28}, 7068 (1983)

\bibitem{greene}
R.~L.~Greene and K.~K.~Bajaj, Phys. Rev. B \textbf{31}, 913 (1985)


\bibitem{yen}
S.~T.~Yen, Phys. Rev. B \textbf{66}, 075340 (2002)


\bibitem{zhilkyun}
A.~G.~Zhilich and B.~K.~Kyuner, Sov. Phys. Semicond. \textbf{15}, 1108 (1981)


\bibitem{per1}
V.~M.~Pereira, V.~N.~Kotov, and A.~H.~Castro Neto, Phys. Rev. B \textbf{78}, 085101 (2008)

\end{thebibliography}
\end{document}